\documentclass[acmtog, screen, nonacm]{acmart}

\usepackage{mwe}
\usepackage{wrapfig}
\usepackage{graphicx}
\usepackage{stackengine}
\usepackage{algorithmic}
\usepackage{algorithm}
\usepackage{subcaption}

\setcopyright{none}

\makeatletter
\let\@authorsaddresses\@empty
\makeatother

\begin{document}

\title{DDFKs: Fluid Simulation with Dynamic Divergence-Free Kernels}

\author{Jingrui Xing$^{1}$}
\author{Yizao Tang$^{2}$}
\author{Mengyu Chu$^{3*}$}
\author{Baoquan Chen$^{3*}$}

\renewcommand{\shortauthors}{Jingrui Xing, et al.}
\newcommand\blfootnote[1]{%
  \begingroup
  \renewcommand\thefootnote{}\footnote{#1}%
  \addtocounter{footnote}{-1}%
  \endgroup
}

\begin{abstract}

\blfootnote{* indicates corresponding authors. \\Affiliations: $^1$School of Intelligence Science and Technology, Peking University, $^2$School of Electronics Engineering and Computer Science, Peking University, $^3$Peking University}
Fluid simulations based on memory-efficient spatial representations like implicit neural spatial representations (INSRs) and Gaussian spatial representation (GSR), where the velocity fields are parameterized by neural networks or weighted Gaussian functions, has been an emerging research area. %
Though advantages over traditional discretizations like spatial adaptivity and continuous differentiability of these spatial representations are leveraged by fluid solvers, solving the time-dependent PDEs that governs the fluid dynamics remain challenging, especially in incompressible fluids where the divergence-free constraint is enforced. %
In this paper, we propose a grid-free solver Dynamic Divergence-Free Kernels (DDFKs) for incompressible flows based on divergence-free kernels (DFKs). Each DFK is incorporated with a matrix-valued radial basis function and a vector-valued weight, yielding a divergence-free vector field. We model the continuous flow velocity as the sum of multiple DFKs, thus enforcing incompressibility while being able to preserve different level of details. %
Quantitative and qualitative results show that our method achieves comparable accuracy, robustness, ability to preserve vortices, time and memory efficiency and generality across diverse phenomena to state-of-the-art methods using memory-efficient spatial representations, while excels at maintaining incompressibility. Though our first-order solver are slower than fluid solvers with traditional discretizations, our approach exhibits significantly lower numerical dissipation due to reduced discretization error. We demonstrate our method on diverse incompressible flow examples with rich vortices and various solid boundary conditions. %
\end{abstract}

\begin{teaserfigure}
    \centering
    \newcommand{\formattedgraphics}[1]{\includegraphics[trim=0 0 0 0,clip,width=0.21\textwidth]{#1}}
    \formattedgraphics{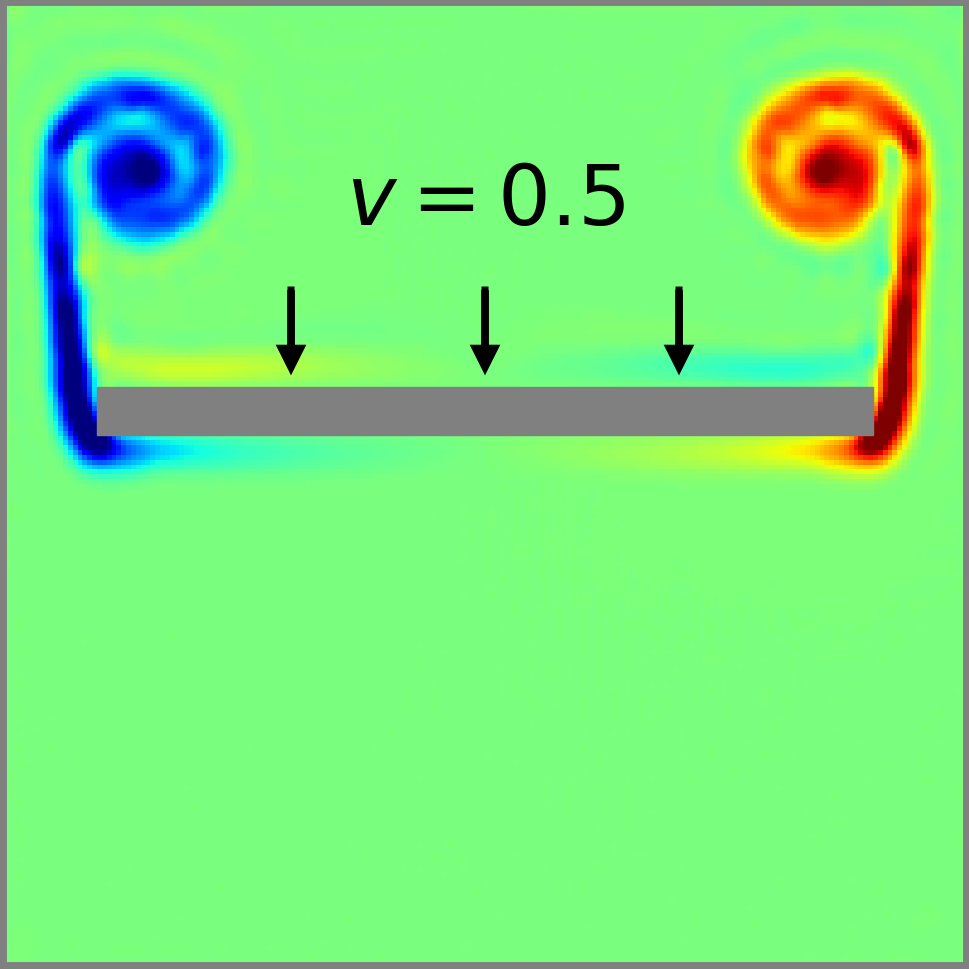}
    \hspace{-4.7pt}
    \formattedgraphics{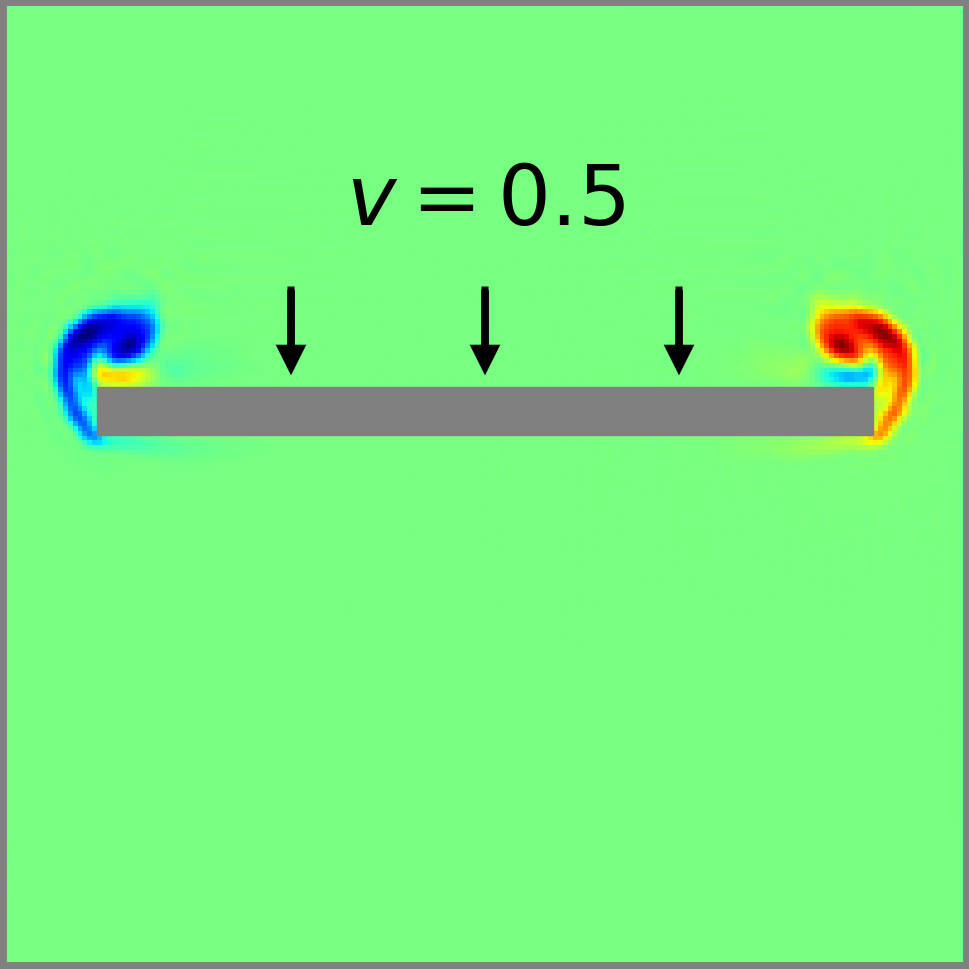}
    \hspace{-4.7pt}
    \formattedgraphics{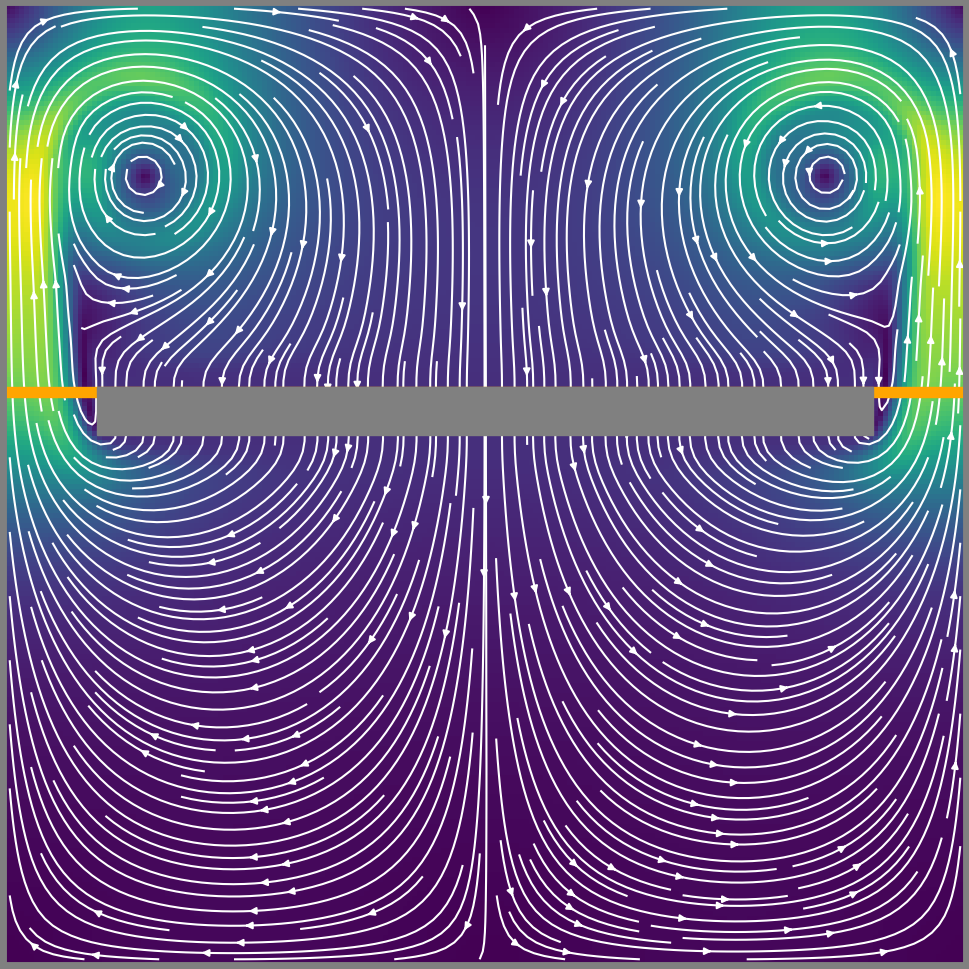}
    \hspace{-4.7pt}
    \formattedgraphics{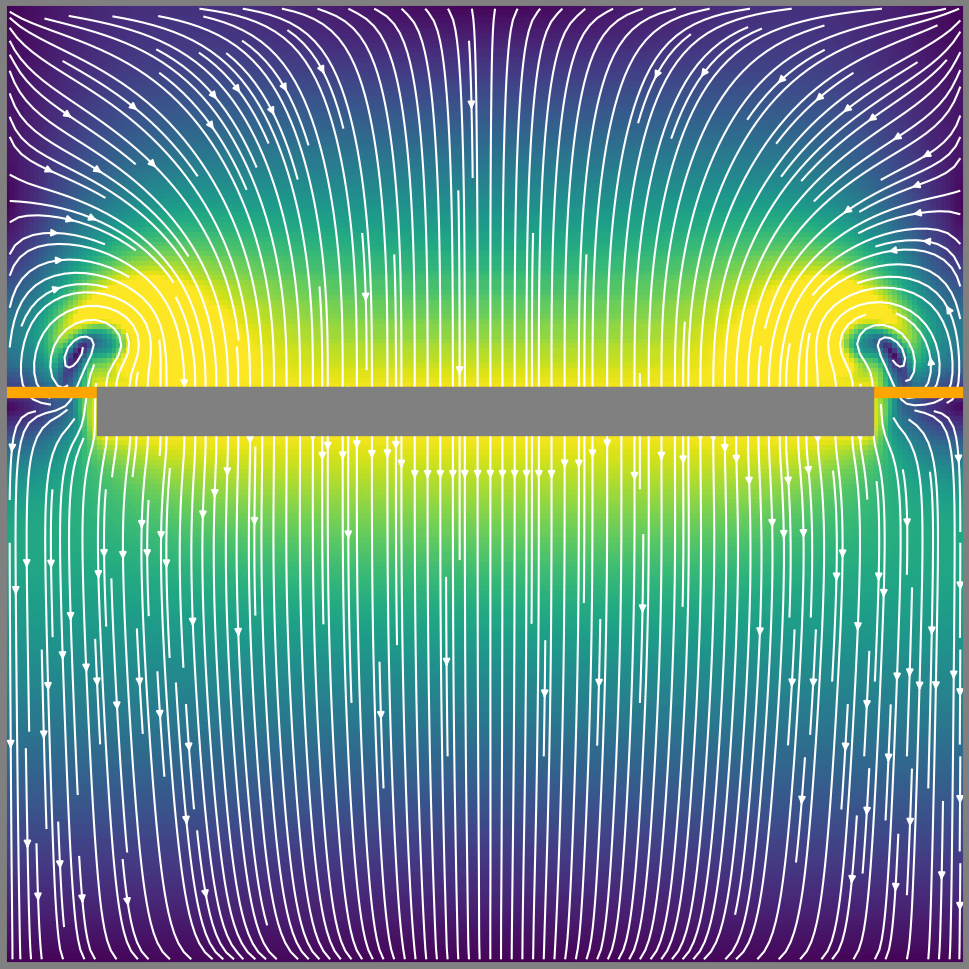}
    \hspace{-4.7pt}
    \includegraphics[trim=0 0 0 0,clip,width=0.156\textwidth]{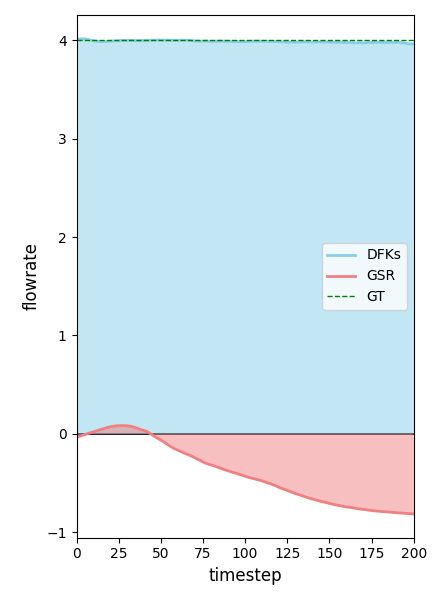}
    \caption{A piston is moving downwards at a constant speed inside a closed vessel, simulated with divergence-free kernels (ours) and GSR, respectively. Left two figures: the vorticity field at frame 50 of our simulation and GSR's, respectively. The third and fourth figure: visualizations of the velocity field at frame 50 of ours and GSR's, respectively, with streamlines plotted along the velocity field and background color reflecting the magnitude of velocity. Right: flowrate measured on the orange lines on the third and fourth figure at each timestep, with upward being positive. The flowrate is calculated by integrating the $y$-component of the velocity along the orange lines.}
    \label{fig:teaser}
\end{teaserfigure}

\maketitle

\section{Introduction}

Fluid simulations remain a persistent challenge in computer graphics, requiring expressive representations to capture rich spatial details while supporting efficient, accurate temporal evolution for modeling nonlinear and chaotic dynamics. Traditional fluid solvers adopt grid-based or particle-based spatial discretizations, corresponding to Eulerian and Lagrangian frameworks, with hybrid approaches also being widely proposed. Such straightforward spatial discretizations facilitate efficient computation for resolving temporal fluid dynamics.

Despite substantial progresses achieved by these methods, curse of dimensionality can easily occur in traditional spatial discretizations in terms of representing fine structures of flow fields. To fundamentally solve this issue, memory-efficient spatial representations are proposed as novel approaches for parameterizing flow fields, leading an emerging trend in fluid simulation. \citet{chen2023insr} and \citet{jian2023nmcf} use neural networks as implicit neural spatial representations (INSRs) to replace explicit data structures like grid or particles. These methods leverage the continuous differentiability of a neural network to solve PDEs, achieving fluid simulations with spatial adaptivity. Furthermore, \citet{xing2025gaussian-fluids} propose Gaussian fluids featuring the Gaussian spatial representation (GSR), where a velocity field is represented with multiple weighted Gaussian functions. This approach achieves comparable spatial adaptivity and memory compactness as INSRs, while significantly improves efficiency on calculating the spatial differentials and vorticity fidelity with high stability.

With novel memory-efficient spatial representations being proposed, maintaining divergence-free velocity fields throughout the simulations becomes challenging.
Fluid solvers based on INSRs adopt the operator-splitting technique typically used in traditional Eulerian methods, where the Navier-Stokes equations are divided into an advection and a projection step. While the advection step is solved by learning a neural network that fits the advected velocity field in both methods, different approaches for projection are adopted in \cite{chen2023insr} and \cite{jian2023nmcf}. The former method directly optimizes a PDE-loss derived from the projection step to learn a neural network that represents the pressure field, which requires calculating the second derivative of a neural network, leading to low computational efficiency. The latter leverages the walk-on-stars (WoSt) to estimate the pressure gradient on each queried point, which makes the simulation unstable due to the high variance of Monte Carlo methods with limited samples.
\citet{xing2025gaussian-fluids} avoid the pressure projection step by solving the vorticity form of the Navier-Stokes equation. However, a physics-informed divergence loss is still required to make the fluid incompressible, which introduces contradiction in gradients to the optimization process, reducing the convergence rate. More importantly, the divergence loss cannot be reduced to zero entirely even after a sufficient number of iterations, which results in severe compression of the fluid under certain extreme scenarios, as illustrated in Figure~\ref{fig:teaser}.

In this paper, we propose a grid-free fluid solver based on the dynamic divergence-free kernels (DDFKs), which perfectly resolves the challenge of maintaining divergence-free velocity fields throughout the entire simulation, while retaining all the advantages of memory-efficient spatial representation methods. The contributions of our method are as follows:
\begin{itemize}
    \item introducing the representation DFKs to forward simulation for fluids;
    \item a novel approach to keep the fluid unconditionally incompressible;
    \item a grid-free solver featuring the DDFKs to simulate intricate vortical flow phenomena with diverse solid boundaries.
\end{itemize}

\section{Related Work}

\paragraph{Traditional Fluid Simulation} Since Stable Fluids \cite{10.1145/311535.311548} established the foundation of fluid simulation method in computer graphics by introducing the Chorin’s projection method \cite{chorin1968numerical} and semi-Lagrangian advection scheme \cite{robert1981stable, sawyer1963semi}, researchers have developed a lot of works to improve the efficiency and accuracy under the Eulerian framework. They accelerate the pressure projection step by using sparse adaptive grid structures \cite{aanjaneya2017power, setaluri2014spgrid, mcadams2010parallel} and develop more accurate advection methods for reducing numerical dissipation \cite{selle2008unconditionally, kim2005flowfixer, zehnder2018advection}.
There are also works placing quantities other than the velocity field on the grid for advection, such as impulse \cite{feng2022impulse}, covector \cite{nabizadeh2022covector}, and vorticity \cite{liu2001numerical,wang2024eulerian}. These methods lead to better vorticity preservation and demonstrate improved stability in capturing fine-scale features.

Smoothed Particle Hydrodynamics (SPH) method was introduced to computer graphics \citet{muller2003particle}, led to a large body of research into enhancing incompressibility and stability\cite{ihmsen2013implicit,he2025semi,bender2015divergence}, improving spatial adaptivity\cite{adams2007adaptively} and mitigating errors due to interpolation inconsistencies in sparsely sampled regions\cite{band2018mls, westhofen2023comparison}. A comprehensive overview of SPH techniques can be found in \citet{koschier2022survey}.
The position-based fluids (PBF) method\cite{macklin2013position, diaz2025implicit} offers exceptional stability with larger time steps by converting SPH into a constraint formulation. Vortex methods typically use Lagrangian representations such as particles\cite{sella2005vortex}, filaments\cite{weissmann2010filament}, and sheets\cite{pfaff2012lagrangian}. They reformulate the fluid equations into vorticity-velocity form and exhibit difficulties in geometric and boundary treatments. 

In addition to pure Eulerian or Lagrangian approaches, there exist hybrid methods that combine both frameworks\cite{harlow1964particle, brackbill1986flip, jiang2015affine, tang2025granule}, as well as techniques that integrate vorticity and velocity formulations\cite{koumoutsakos2008flow, wang2025vpfm}. Conventional fluid simulation methods typically struggle with limitations inherent in their spatial discretization schemes, prompting various enhancements throughout the years. Our research follows the concept adopting continuous representations that provide greater expressive capability. While optimization-driven methodology may not match the performance of current state-of-the-art techniques that rely on explicit representations in certain aspects, it demonstrates superior capabilities in preserving fine details and adapting to spatial variations. We believe this represents a valuable and promising avenue for future investigation in fluid simulation.

\paragraph{Memory-Efficient Spatial Representations for Fluid Simulation}Another emerging direction involves the integration of Implicit Neural Representations (INR) and Gaussian functions within simulation frameworks. Many studies have adopted these representations as mass distribution functions, combining them with established numerical solvers—such as Material Point Methods—to enable physics-based scene manipulation and animation \cite{feng2024pienerfphysicsbasedinteractiveelastodynamics, xie2024physgaussianphysicsintegrated3dgaussians}. For instance, \citet{deng2023nfm} employs INR to store flow maps and achieves non-dissipative simulation results through a grid-based fluid solver with extended advection capabilities. Nevertheless, relatively limited research has investigated the application of these innovative representations as spatial functions for solving time-dependent partial differential equations (PDEs). Among the most pertinent works to our research are Implicit Neural Spatial Representations for PDEs (INSR) \cite{chen2023insr} and Neural Monte Carlo Fluids (NMC) \cite{jian2023nmcf}. The INSR framework utilizes INR as spatial representations for both fluid and soft-body simulations, addressing temporal PDEs through optimization techniques. In contrast, the NMC approach capitalizes on the inherent continuity properties of INR by implementing the walk-on-spheres method for pressure computation and enhancing the INR architecture to accommodate boundary conditions effectively. Most recently, \citet{xing2025gaussian-fluids} presents a grid-free fluid solver based on Gaussian Spatial Representation (GSR), modeling the continuous fluid velocity field as a weighted sum of multiple Gaussian functions. it offers an alternative to traditional Eulerian/Lagrangian methods with remarkable accuracy and detail preservation, though it cannot guarantee strict divergence-free conditions.

\section{Background}

The dynamic of fluids is typically characterized by its velocity $\boldsymbol u(\boldsymbol x,t)$, which is a vector field defined on the Cartesian product of the fluids' spatial domain $\Omega$ and time interval of interest $[0, T]$.
For incompressible flows, $\boldsymbol u$ is governed by the Navier-Stokes equations with the divergence-free constraint:
\begin{align}
    \frac{\partial\boldsymbol u}{\partial t}+(\boldsymbol u\cdot\nabla)\boldsymbol u&=-\frac 1\rho\nabla p+\nu\nabla^2\boldsymbol u+\boldsymbol g,\label{eqn:NS}\\
    \nabla\cdot\boldsymbol u&=0,\label{eqn:div-free}
\end{align}
where $\rho$ is the fluid density, $p$ is pressure, $\nu$ is kinematic viscosity, and $\boldsymbol g$ is acceleration due to external force.
If the fluid is inviscid (i.e. $\nu=0$), assuming $\rho$ is a constant, taking the curl of Eq.~\ref{eqn:NS} yields the governing equation of the vorticity $\boldsymbol\omega=\nabla\times\boldsymbol u$ as
\begin{equation}
    \frac{\partial\boldsymbol\omega}{\partial t}+(\boldsymbol u\cdot\nabla)\boldsymbol\omega=(\boldsymbol\omega\cdot\nabla)\boldsymbol u+\nabla\times\boldsymbol g.\label{eqn:NS-vor}
\end{equation}

Given the initial state of the fluid
\begin{align}
    \boldsymbol u(\boldsymbol x,0)=\boldsymbol u^0(\boldsymbol x)\quad\forall\boldsymbol x\in\Omega
\end{align}
and the boundary conditions
\begin{align}
    \boldsymbol u(\boldsymbol x,t)=\boldsymbol u_\mathrm b(\boldsymbol x,t)\quad&\forall\boldsymbol x\in\Gamma_1,\label{eqn:no-slip}\\
    \boldsymbol u(\boldsymbol x,t)\cdot\boldsymbol n_{\Gamma_2}(\boldsymbol x)=f_\mathrm b(\boldsymbol x,t)\quad&\forall\boldsymbol x\in\Gamma_2,\label{eqn:free-slip}
\end{align}
where $\boldsymbol u^0,\boldsymbol u_\mathrm b,f_\mathrm b$ are known functions, $\Gamma_1,\Gamma_2\subseteq\partial\Omega$ are sets where the no-slip and free-slip boundary conditions apply, respectively, $\boldsymbol n_{\Gamma_2}$ is the normal of the boundary geometry, numerical solvers typically discretize the time interval of interest into multiple timesteps, and calculate the evolution of flow velocity in each timestep from $\boldsymbol u^0$ following Eq.~\ref{eqn:NS}or Eq.~\ref{eqn:NS-vor}, while ensuring the divergence-free constraint and the boundary conditions are satisfied at each timestep.

\section{Divergence-Free Kernels}

Our solver leverages the divergence-free kernels (DFKs) to parameterize the velocity fields of an incompressible flow, while leaving the classical temporal discretization unchanged. In this section, we present the mathematical formula of a DFK and some essential ingredients for the solver to evolve the DFKs.

\paragraph{Mathematical Formulae}
Given the spacial dimention $d$, the $i$-th DFK is a particle with several attributes: Its position $\boldsymbol p_i\in\mathbb R^d$, kernel radius $h_i\in\mathbb R$, and weight $\boldsymbol w_i\in\mathbb R^d$.
We first construct a scalar-valued kernel with a radial basis function $\phi(r)$ as
\begin{equation}
    \phi_i(\boldsymbol x)=\phi\left(\frac{\Vert\boldsymbol x-\boldsymbol p_i\Vert}{h_i}\right).
\end{equation}
Then a matrix-valued kernel $\boldsymbol\psi_i:\mathbb R^d\to\mathbb R^{d\times d}$ can be further derived from $\phi_i$:
\begin{equation}
    \boldsymbol\psi_i(\boldsymbol x)=(-\mathbf I\nabla\cdot\nabla+\nabla\nabla^\top)\phi_i(\boldsymbol x),
\end{equation}
where $\mathbf I$ is the identity matrix and $\nabla\nabla^\top$ is the Hessian operator. The vector field induced by the DFK is defined as
\begin{equation}
    \tilde{\boldsymbol u}_i(\boldsymbol x)=\boldsymbol\psi_i(\boldsymbol x)\boldsymbol w_i.
\end{equation}
It can be shown that $\nabla\cdot\tilde{\boldsymbol u}_i=0$ for any given weight $\boldsymbol w_i$ as long as the radial basis function $\phi(r)$ is $\mathcal C^2$-continuous \cite{Ni2025DFK}. Hence by combining $N$ DFKs, we obtain a representation of a divergence-free vector field
\begin{equation}
    \tilde{\boldsymbol u}(\boldsymbol x)=\sum_{i=1}^N\tilde{\boldsymbol u}_i(\boldsymbol x)=\sum_{i=1}^N\boldsymbol\psi_i(\boldsymbol x)\boldsymbol w_i\label{eqn:DFKs-definition}
\end{equation}
with learnable parameters $\Theta=\{\boldsymbol p_i,h_i,\boldsymbol w_i:i=1,\cdots,N\}$. We choose the radial basis function as the $\mathcal C^4$-continuous piecewise-polynomial function proposed in \cite{Wendland1995}:
\begin{equation}
    \phi(r)=\begin{cases}
        (1-r)^6(35r^2+18r+3),&r\in[0,1]\\
        0,&\text{otherwise}
    \end{cases},\label{eqn:Wen4-rbf}
\end{equation}
\begin{wrapfigure}{r}{0.15\textwidth}
    \vspace{-10pt}
	\centering
	\includegraphics[width=0.99\linewidth,trim={0 0 0 0},clip]{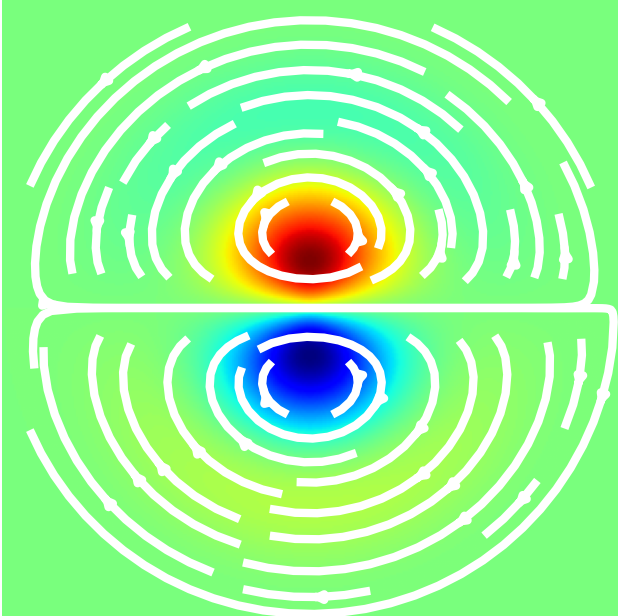}
\end{wrapfigure}
since it is proven the optimal choice for DFKs in \cite{Ni2025DFK}.
The inset figure displays the streamline plot of the velocity field induced by a single DFK in 2D when taking this formula, with the background color reflecting its vorticity.
Substituting Eq.~\ref{eqn:Wen4-rbf} into Eq.~\ref{eqn:DFKs-definition} yields the final formula of the DFKs:
\begin{equation}
    \tilde{\boldsymbol u}(\boldsymbol x)=\sum_{i=1}^Nf(r)\boldsymbol w_i+g(r)(\boldsymbol w_i\cdot\boldsymbol y)\boldsymbol y,\label{eqn:DFKs}
\end{equation}
where $\boldsymbol y=\frac{\boldsymbol x-\boldsymbol p_i}{h_i}$, $r=\Vert\boldsymbol y\Vert$,
\begin{equation}
    \begin{aligned}
        &f(r)=56(1-r)^4[-5(d+5)r^2+(d-1)(4r+1)],\\
        &g(r)=1680(1-r)^4.
    \end{aligned}
\end{equation}

\paragraph{Computational Savings}
Evaluating the vector field through Eq.~\ref{eqn:DFKs} requires $O(N)$ floating-point operations, which is computationally overwhelming for tasks that include frequent queries. To reduce the time complexity, we take advantage of the compactness of each DFK's non-zero domain by adopting a hash grid which is typically used in particle-based fluid simulations to accelerate the neighbor search, reducing the floating-point operations to $O(1)$ on a single query.

\paragraph{Advantages on Spatial Differentials}
The gradient fo the DFKs can be obtained by taking derivatives w.r.t. $\boldsymbol x$ on both side of Eq.~\ref{eqn:DFKs}:
\begin{equation}\small
    \begin{aligned}
    &\nabla\tilde{\boldsymbol u}(\boldsymbol x)=\sum_{i=1}^N\\
    &\frac 1{h_i}\left(\frac{f'(r)}{r}\boldsymbol w_i\boldsymbol y^\top+\frac{g'(r)}{r}(\boldsymbol w_i\cdot\boldsymbol y)\boldsymbol y\boldsymbol y^\top+g(r)\boldsymbol y\boldsymbol w_i^\top+g(r)(\boldsymbol w_i\cdot\boldsymbol y)\mathbf I\right),
    \end{aligned}
\end{equation}
from which we can further derive the divergence and curl operators.
Unlike implicit neural representations that rely on auto-differentiation for computing differential quantities, DFKs allow direct and efficient differentiation, with the same time complexity as evaluating the DFKs field itself.
Furthermore, the $C^4$-continuous property of $\phi(r)$ guarantees that $\nabla\tilde{\boldsymbol u}$ has $C^1$-continuity, which does not introduce numerical error due to discontinuity like the clamped Gaussian functions in the GSR.
These inherent advantages leads to faster and more accurate optimization when enforcing physics-based constraints, making DFKs a computationally superior choice for dynamic simulations.

\section{Algorithm}
In this section, we will introduce our Dynamic Divergence Free Kernels (DDFK) solver to simulate the inviscid incompressible flows with zero external force using DFKs as representation. The algorithm of our method is mainly composed of three parts: initialization, physics-based optimization, and reinitialization. Algorithm \ref{alg:fluid-solver} is a brief overview of our method.

\begin{algorithm}
\caption{Fluid solver with DDFK}
\label{alg:fluid-solver}
\begin{algorithmic}[1]
\STATE $\tilde{\boldsymbol u}^0 \leftarrow \text{Initialize}(\boldsymbol u^0)$
\FOR{$n \leftarrow 1, \cdots, T$}
\STATE //Reinitialize the DFKs every $N_\mathrm{reinit}$ steps:
\IF{$n\bmod N_\mathrm{reinit}=0$}
\STATE $\tilde{\boldsymbol u}^\star\leftarrow\text{Reinitialize}(\tilde{\boldsymbol u}^{n-1})$
\ELSE
\STATE $\tilde{\boldsymbol u}^\star\leftarrow\tilde{\boldsymbol u}^{n-1}$
\ENDIF
\STATE //An initial guess for physics-based optimization:
\STATE $\tilde{\boldsymbol u}^* \leftarrow \text{AdvectPositions}(\tilde{\boldsymbol u}^\star, \tilde{\boldsymbol u}^{n-1})$
\STATE //Physics-based optimization:
\STATE $\tilde{\boldsymbol u}^n \leftarrow \text{OptimizeLosses}(\tilde{\boldsymbol u}^*, \tilde{\boldsymbol u}^{n-1})$
\ENDFOR
\end{algorithmic}
\end{algorithm}

\subsection{Initialization}
\label{sec:initialization}

We start the simulation by initializing the DDFK $\tilde{\boldsymbol u}^0$ to fit the given initial condition of the flow velocity $\boldsymbol u^0$. Similar to \cite{Ni2025DFK}, we formulate this process as an optimization problem:
\begin{equation}
    \mathop{\arg\min}_\Theta\frac 1{d\vert\Omega\vert}\int_\Omega\Vert\tilde{\boldsymbol u}^0(\boldsymbol x)-\boldsymbol u^0(\boldsymbol x)\Vert_1\mathrm dV.
\end{equation}
However, directly evaluating the integral is difficult, hence we replace the objective function by a Monte-Carlo estimation of it:
\begin{equation}
    \mathcal L_\mathrm{val}=\frac 1{Qd}\sum_{j=1}^Q\Vert\tilde{\boldsymbol u}^0(\boldsymbol x_j)-\boldsymbol u^0(\boldsymbol x_j)\Vert_1,\label{eqn:loss_val}
\end{equation}
where $\boldsymbol x_1,\cdots,\boldsymbol x_Q$ are uniformly randomly sampled from $\Omega$ in each optimization iteration. To further improve the fitting quality in favor of the time-integration process afterward, we employ an additional gradient loss:
\begin{equation}
    \mathcal L_\mathrm{grad}=\frac 1{Qd^2}\sum_{j=1}^Q\Vert\nabla\tilde{\boldsymbol u}^0(\boldsymbol x_j)-\nabla\boldsymbol u^0(\boldsymbol x_j)\Vert_\mathrm{sum},\label{eqn:loss_grad}
\end{equation}
where $\Vert\cdot\Vert_\mathrm{sum}=\sum_k\sum_l\vert[\cdot]_{kl}\vert$ is the sum of the absolute values of all the matrix's elements. The gradient loss provides a supervision on how the DFKs field changes in space in addition to the value itself, leading to a more accurate fitting to the given field on convergence. The total loss in the initialization process is
\begin{equation}
    \mathcal L_\mathrm{init}=\mathcal L_\mathrm{val}+\mathcal L_\mathrm{grad}.
\end{equation}

In practice, we find fitting error can be extremely high near the outer boundary of $\Omega$ in some examples. This is because the samples in the optimization process can only land on one side of the outer boundary, while the DFKs do not provide an intuitive extrapolation of a field (see paragraph \textit{Zero net momentum} in Section 7 in \cite{Ni2025DFK}) like GSR. We fix this by slightly expanding the fluid domain into $\Omega_\varepsilon$ and constructing an analytic continuation $\hat{\boldsymbol u}^0$ defined on $\Omega_\varepsilon$ of $\boldsymbol u^0$, followed by changing $\Omega$ into $\Omega_\varepsilon$ and $\boldsymbol u^0$ into $\hat{\boldsymbol u}^0$ in Eq.~\ref{eqn:loss_val} and Eq.~\ref{eqn:loss_grad}.

Before optimization, we initialize the positions of the DFKs as the coordinates of the vertices of a grid that covers $\Omega_\varepsilon$. The radii of the DFKs are set to
\begin{equation}
    h_i=\eta\left[\frac{\Gamma(1+d/2)\vert\Omega_\varepsilon\vert}{N\pi^{d/2}}\right]^{1/d}
\end{equation}
in accordance with \cite{Ni2025DFK}, where $\Gamma(\cdot)$ is the Gamma function and $\eta$ is a hyperparameter, while the weights are set to zero. We then use the Adam optimizer with a scheduler which decay the learning rate after certain continual iterations without loss descending to obtain initial DFKs ready for simulation.

\subsection{Physics-Based Optimization}

The time integration is done by optimizing the learnable parameters of the DFKs to minimize a series of physics-based losses. By limiting the search space to the divergence-free vector fields represented by DFKs, our solver avoids physics-informed divergence loss which is commonly used in other methods based on memory-efficient spatial representations like INSRs and GSR, which subsequently leads to a more direct optimization while strictly satisfying the incompressible condition.

\subsubsection{Advection-Based Initial Guess}

To ensure a better convergence result and temporal consistency, we advect the particle positions forward by one time step while leaving other parameters of the DFKs unchanged before the optimization. Specifically, we apply the 4-th order Runge-Kutta convention with velocity field $\tilde{\boldsymbol u}^{n-1}$ to move the particle positions. Note that this step is different from the behavior of typical advection in traditional particle-based methods like SPH or PIC-FLIP, where a particle is moved along its own velocity, while the velocity used to move a DFK is computed with neighboring particles.

Though this step can only provide a very roughly approximated solution of the advection equation $\frac{\partial\boldsymbol u}{\partial t}+(\boldsymbol u\cdot\nabla)\boldsymbol u=\boldsymbol 0$, it is a viable choice for an initial guess for the subsequent optimization. Updating particle positions via the current velocity field facilitates the subsequent optimization in searching for a local optimum that is closer to the true advected state, thereby yielding simulations with overall flow velocities more closely aligned with physical reality. The effectiveness of this advection step is validated through an ablation test in Section~\ref{sec:ablation}.

\subsubsection{The Vorticity Loss}

Since the DFKs cannot represent a vector field with non-zero divergence, the operator splitting method which divides Eq.~\ref{eqn:NS} into advection and projection steps typically used in some traditional solvers is no longer valid in our case. Therefore, we borrow the idea from vortex methods and solve Eq.~\ref{eqn:NS-vor} instead by optimizing the DFKs to minimize the following vorticity loss:
\begin{equation}
    \mathcal L_\mathrm{vor}=\frac 1{Q\hat d}\sum_{j=1}^Q\Vert\nabla\times\tilde{\boldsymbol u}^n(\boldsymbol x_j)-\omega(\boldsymbol x_j)\Vert_1,
\end{equation}
where $\hat d=1$ in 2D and $\hat d=3$ in 3D since the vorticity is a scalar in 2D and a vector in 3D, $\boldsymbol x_1,\cdots,\boldsymbol x_Q$ are uniformly randomly sampled from $\Omega$ in each iteration, $\omega(\boldsymbol x)$ is the vorticity field advected from $\nabla\times\tilde{\boldsymbol u}^{n-1}(\boldsymbol x)$.
Denote $\boldsymbol\Phi^{n-1}:\mathbb R^d\to\mathbb R^d$ as the mapping from a point at timestep $n-1$ to its position after being transported forward by the RK4 time integration with velocity field $\tilde{\boldsymbol u}^{n-1}(\boldsymbol x)$, and $\boldsymbol\Psi^{n-1}:\mathbb R^d\to\mathbb R^d$ as the inverse mapping of $\boldsymbol\Phi^{n-1}$.
In 2D, $(\boldsymbol\omega\cdot\nabla)\boldsymbol u=\boldsymbol 0$, hence Eq.~\ref{eqn:NS-vor} becomes $\frac{\mathrm D\boldsymbol\omega}{\mathrm Dt}=\boldsymbol 0$, meaning that the vorticity simply transports along the velocity field without evolving:
\begin{equation}
    \omega(\boldsymbol x)=\nabla\times\tilde{\boldsymbol u}^{n-1}(\boldsymbol\Psi^{n-1}(\boldsymbol x)).
\end{equation}
In 3D, the vorticity field evolves according to $\frac{\mathrm D\omega}{\mathrm Dt}=(\nabla\boldsymbol u)\omega$, where the R.H.S. is a matrix product. Hence the vorticity is a line element as mentioned in \cite{wang2024eulerian}, which can be advected under the flow map $\boldsymbol\Phi^{n-1}$ with
\begin{equation}
    \omega(\boldsymbol x)=[\mathrm d\boldsymbol\Psi^{n-1}(\boldsymbol x)]^{-1}\nabla\times\tilde{\boldsymbol u}^{n-1}(\boldsymbol\Psi^{n-1}(\boldsymbol x)),
\end{equation}
where $\mathrm d\boldsymbol\Psi^{n-1}$ is the Jacobian matrix of $\boldsymbol\Psi^{n-1}$.

\subsubsection{Boundary Handling}

We handle the boundary conditions by introducing boundary losses to the optimization system. The no-slip boundary formulated as Eq.~\ref{eqn:no-slip} is enforced with the type-1 boundary loss
\begin{equation}
    \mathcal L_\mathrm{b1}=\frac 1{Q_\mathrm{b1}d}\sum_{j=1}^{Q_\mathrm{b1}}\Vert\tilde{\boldsymbol u}^n(\boldsymbol z_j)-\boldsymbol u_\mathrm b(\boldsymbol z_j,t_n)\Vert_1,
\end{equation}
while the free-slip boundary formulated as Eq.~\ref{eqn:free-slip} is enforced with the type-2 boundary loss
\begin{equation}
    \mathcal L_\mathrm{b2}=\frac 1{Q_\mathrm{b2}}\sum_{j=Q_\mathrm{b1}+1}^{Q_\mathrm{b1}+Q_\mathrm{b2}}\vert\tilde{\boldsymbol u}^n(\boldsymbol z_j)\cdot\boldsymbol n_{\Gamma_2}(\boldsymbol z_j)-f_\mathrm b(\boldsymbol z_j,t_n)\vert,
\end{equation}
where $\boldsymbol z_1,\cdots,\boldsymbol z_{Q_\mathrm{b1}}$ are uniformly randomly sampled from the boundary geometry $\Gamma_1$ where the no-slip condition apply, $\boldsymbol z_{Q_\mathrm{b1}+1},\cdots,\boldsymbol z_{Q_\mathrm{b1}+Q_\mathrm{b2}}$ are sampled from $\Gamma_2$ where the free-slip condition apply, $t_n$ denotes the time at step $n$.

The total loss for the time integration is the linear combination of the vorticity loss and the boundary losses:
\begin{equation}
    \mathcal L_\mathrm{adv}=\mathcal L_\mathrm{vor}+\lambda_\mathrm{b1}\mathcal L_\mathrm{b1}+\lambda_\mathrm{b2}\mathcal L_\mathrm{b2}.
\end{equation}
We then use the same optimizer as the initialization step to minimize $\mathcal L_\mathrm{adv}$ after the particle advection, yielding the DFKs at the next step $\tilde{\boldsymbol u}^n$.

\subsection{Reinitialization}

As the simulation progresses through multiple time steps, the optimization processes can potentially drive DFK particles to cluster in certain local regions, which in turn gives rise to voids where the particle distribution becomes sparse.
To alleviate this issue, we reinitialize the DFKs at the beginning of the timestep every $N_\mathrm{reinit}$ steps, which involves re-learning the current velocity field from scratch and introducing extra particles in specific regions to preserve flow details.

In the re-learning stage, we conduct the same fitting optimization as described in Section~\ref{sec:initialization}, with the target velocity field changed into the DFKs at the last timestep $\tilde{\boldsymbol u}^{n-1}(\boldsymbol x)$. Given that the velocity field typically gains complexity over time, a simple re-learning procedure may be insufficient to resolve the fine-scale structures in $\tilde{\boldsymbol u}^{n-1}(\boldsymbol x)$. Consequently, we borrow some particles from the preceding state $\tilde{\boldsymbol u}^{n-1}$ to supplement the under-fitting areas. Specifically, we evaluate the gradient error at the position of each preceding DFK:
\begin{equation}
    \boldsymbol G^\mathrm{err}_i=\nabla\tilde{\boldsymbol u}^\star(\boldsymbol p^{n-1}_i)-\nabla\tilde{\boldsymbol u}^{n-1}(\boldsymbol p^{n-1}_i),
\end{equation}
where $\boldsymbol p^k_i$ denotes the position of the $i$-th particle of the DFK state $\tilde{\boldsymbol u}^k$. Next, we append the particles which satisfy $\Vert\boldsymbol G^\mathrm{err}_i\Vert_\mathrm{sum}>\alpha\cdot\mathrm{mean}\{\Vert\boldsymbol G^\mathrm{err}_j\Vert_\mathrm{sum}:j=1,\cdots,N\}$ to $\tilde{\boldsymbol u}^\star$. Finally, we conduct another optimization for $\tilde{\boldsymbol u}^\star$ to fit $\tilde{\boldsymbol u}^{n-1}(\boldsymbol x)$. We choose $\alpha=2$ in all the examples in this paper.

\section{Results}

We evaluate our method on various examples across different initial conditions with diverse boundary shapes and types. In contrast to most solvers relying on first-order optimization, our approach exhibits comparable learning rate robustness to GSR, with neither requiring hyperparameter tuning across varying scenarios. To accommodate test cases of varying scales, we normalize the entire fluid domain and boundary geometries to a canonical size, with corresponding adjustments to the initial velocity field. Through this scaling strategy, a single set of parameter configurations is sufficient for all 2D test cases, and a separate configuration for all 3D test cases.

We first apply our method on an example with analytical solution to quantify its numerical accuracy and convergence rate. We then validate its effectiveness through comparative experiments with traditional solvers and other memory-efficient spatial representations, which highlights its reduced numerical dissipation and superior ability to satisfy the incompressible condition. Moreover, we demonstrate that DFKs are capable of handling complex dynamics with highly intricate fluid simulations. Finally, a series of ablation studies are conducted to assess the contribution of each key component in our method. The performances for all examples are provided at the end of this section.

\subsection{Quantitative Study}

We conduct a quantitative study using Taylor Green vortex experiment, analyzing the numerical accuracy of our method and the convergence rates of the optimizations during initialization and time integration. In this example, the initial velocity field is set to
\begin{equation}
    \boldsymbol u(\boldsymbol x,\boldsymbol y)=\begin{bmatrix}
\sin x \cos y \\
-\cos x \sin y
\end{bmatrix}
\end{equation}
defined on the fluid domain $\mathcal{D} = [0, 2\pi] \times [0, 2\pi]$. We then run the
simulation for 100 frames with a 0.001 seconds timestep.

As this velocity field will remain constant on inviscid incompressible fluids, we can measure the numerical error by calculating the mean squared error (MSE) between the simulated velocity fields and initial velocity field $\boldsymbol u(\boldsymbol x,\boldsymbol y)$ sampled on a $60 \times 60$ uniform grid of certain frame.

The results in Table~\ref{tab:error_compare_vertical} show that our method has much lower numerical error than INR-based methods and GSR.

\begin{table}[htbp]
  \centering
  \begin{tabular}{c|ccc}
    \textbf{Method} & Frame 0 & Frame 50 & Frame 100 \\
    \hline
    \textbf{Eulerian} & $2.432 \times 10^{-4}$ & $9.757 \times 10^{-3}$ & $2.019 \times 10^{-2}$ \\
    \hline
    \textbf{INSR} & $8.998 \times 10^{-7}$ & $1.715 \times 10^{-5}$ & $1.992 \times 10^{-5}$ \\
    \hline
    \textbf{NMC} & $1.829 \times 10^{-4}$ & $6.492 \times 10^{-4}$ & $1.725 \times 10^{-3}$ \\
    \hline
    \textbf{GSR} & $9.957 \times 10^{-8}$ & $2.510 \times 10^{-7}$ & $2.181 \times 10^{-7}$ \\
    \hline
    \textbf{Ours} & $2.343 \times 10^{-8}$ & $2.484 \times 10^{-8}$ & $3.305 \times 10^{-8}$ \\
  \end{tabular}
  \caption{Quantitative comparison using the Taylor-Green vortex example. MSE between the simulated velocity fields of different methods and the ground truth on frame 0, 50 and 100.}
  \label{tab:error_compare_vertical}
\end{table}

\subsection{Validation}

\begin{figure*}[htbp]\small
\centering
\newcommand{\formattedgraphics}[1]{\includegraphics[trim=110 50 170 65,clip,width=0.175\textwidth]{#1}}
\newcommand{\formattedgraphicslst}[1]{\includegraphics[trim=100 50 144 65,clip,width=0.21\textwidth]{#1}}
\newcommand{\negspace}{\hspace{-1.8pt}}
\newcommand{\negvspace}{\vspace{-6pt}}
\begin{tabular}{p{2.5cm} l}  %
  \textbf{Semi-Implicit Euler\newline $512\times 512$ \newline ($2$ MB)} & 
  \begin{minipage}{0.8\textwidth}\negvspace
    \formattedgraphics{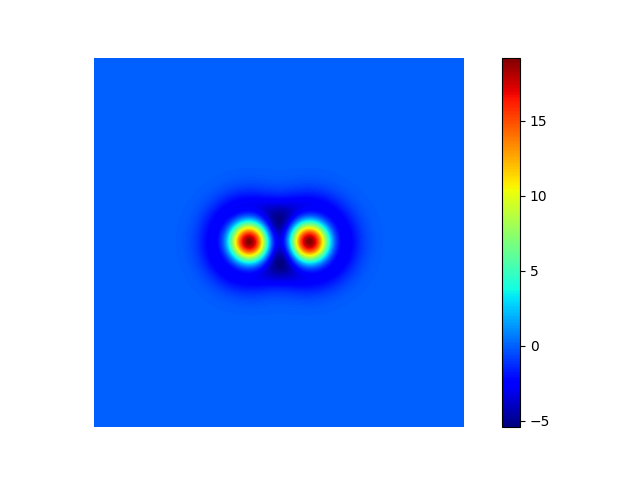}
    \negspace
    \formattedgraphics{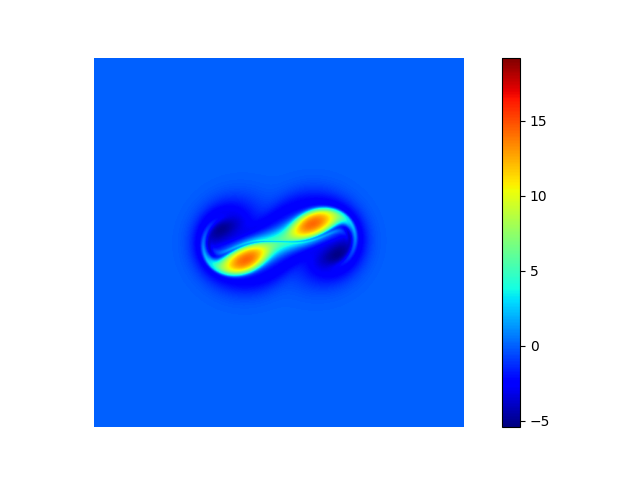}
    \negspace
    \formattedgraphics{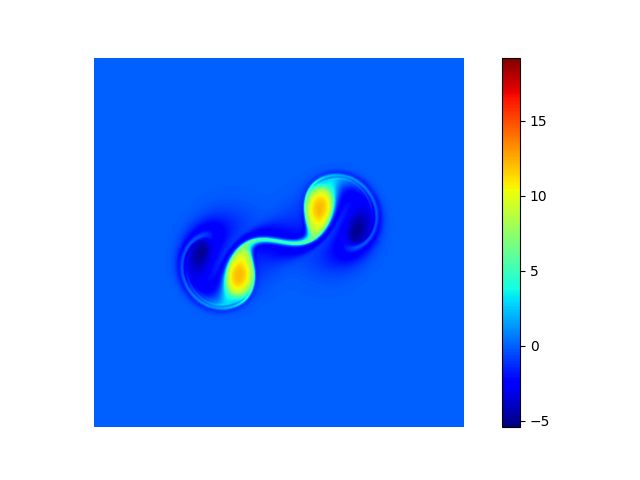}
    \negspace
    \formattedgraphics{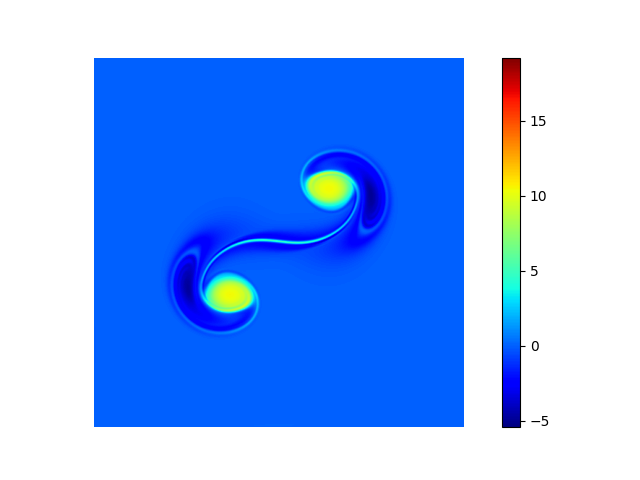}
    \negspace
    \formattedgraphicslst{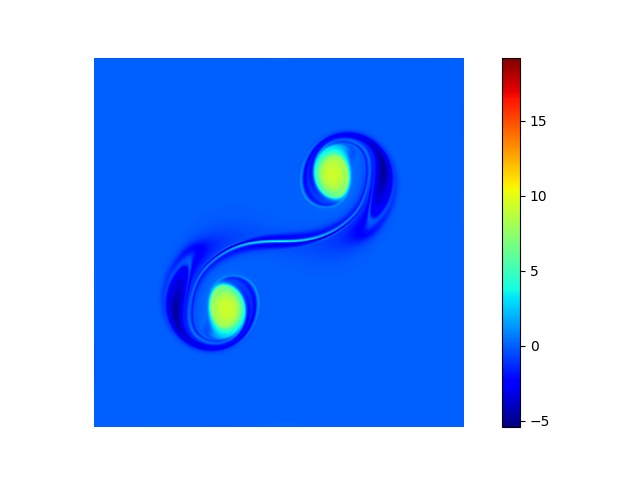}
  \end{minipage} \\[-6pt]
  \textbf{INSR \newline \cite{chen2023implicit} \newline ($32.1$ KB)} & 
  \begin{minipage}{0.8\textwidth}\negvspace
    \formattedgraphics{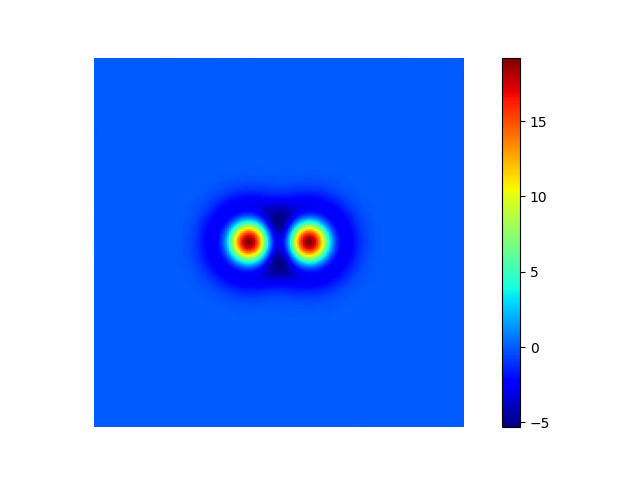}
    \negspace
    \formattedgraphics{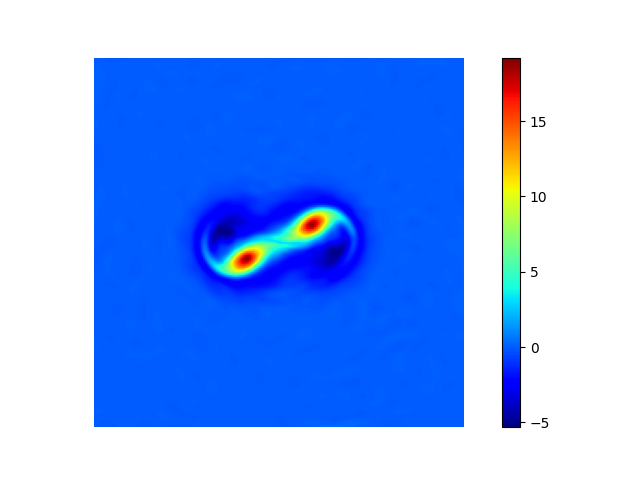}
    \negspace
    \formattedgraphics{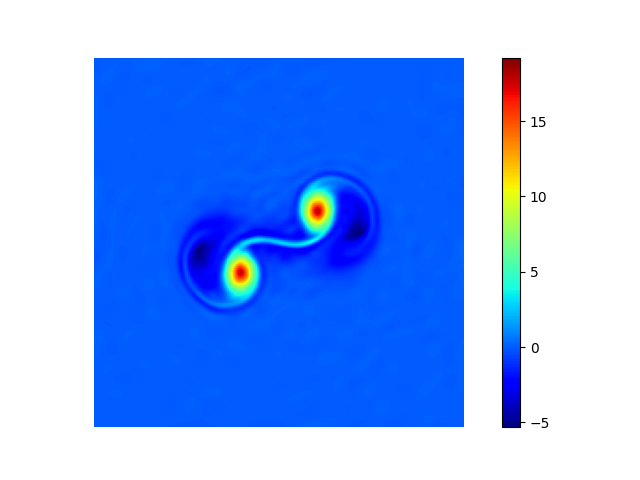}
    \negspace
    \formattedgraphics{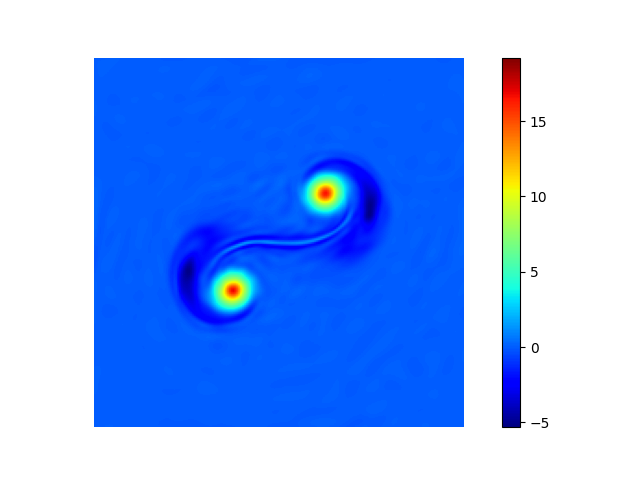}
    \negspace
    \formattedgraphicslst{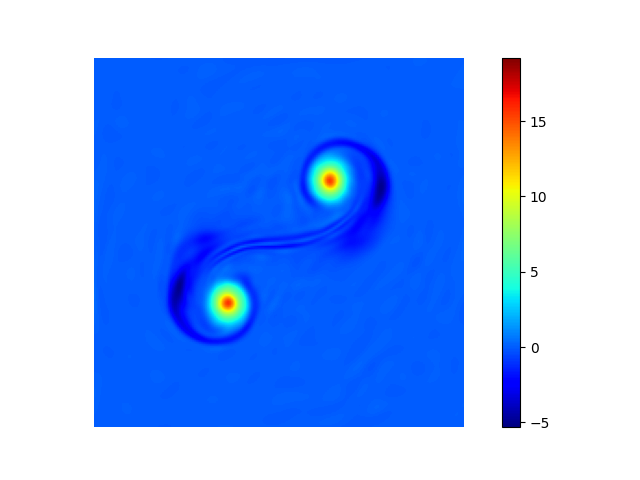}
  \end{minipage} \\[-6pt]
  \textbf{GSR\newline
  \cite{xing2025gaussian-fluids}
  \newline $5041\sim 5511$ particles \newline ($148.5$ KB)} & 
  \begin{minipage}{0.8\textwidth}\negvspace
    \formattedgraphics{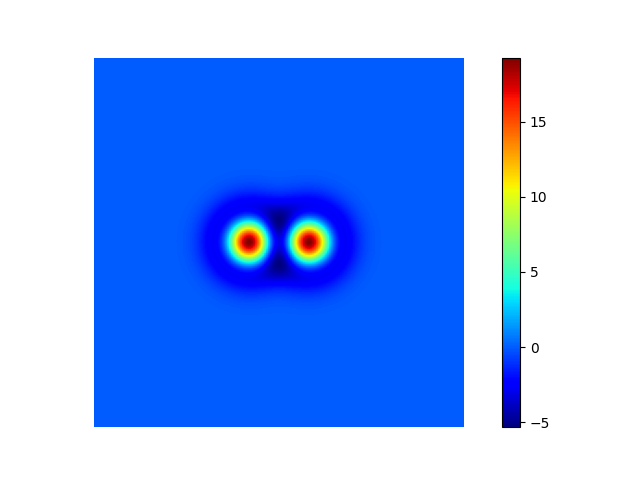}
    \negspace
    \formattedgraphics{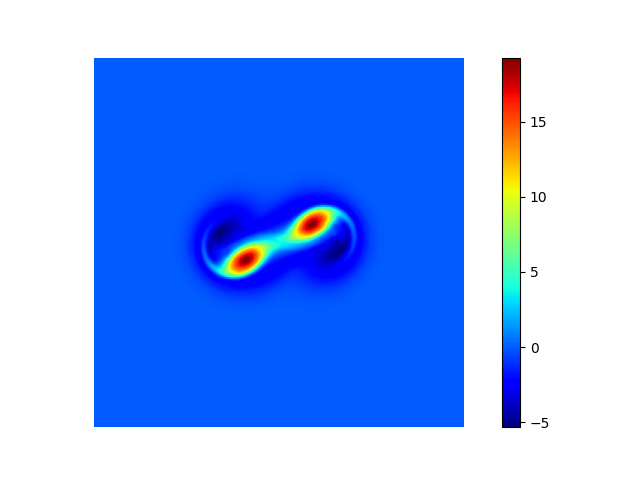}
    \negspace
    \formattedgraphics{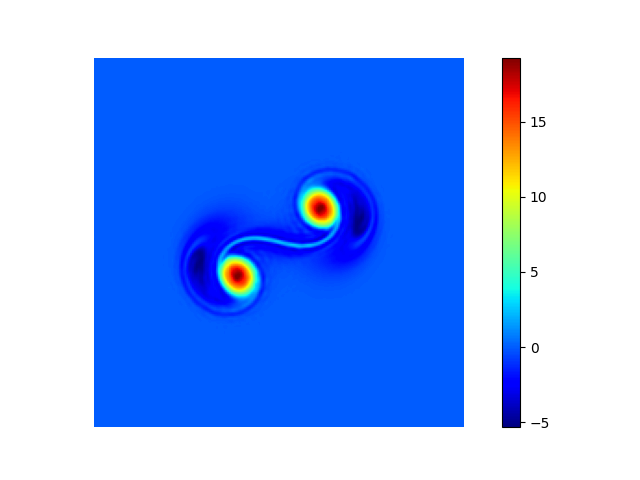}
    \negspace
    \formattedgraphics{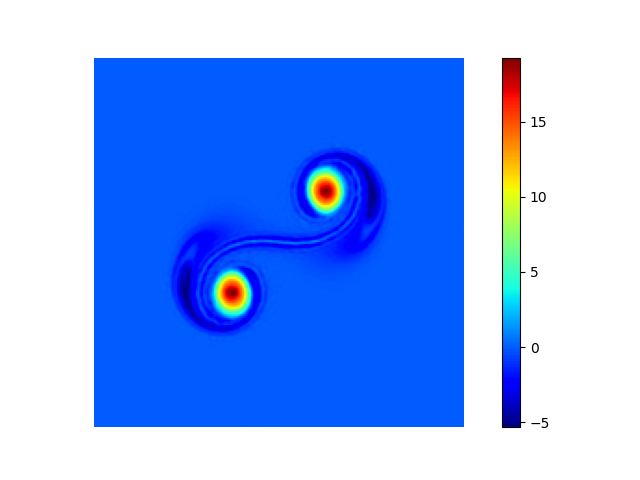}
    \negspace
    \formattedgraphicslst{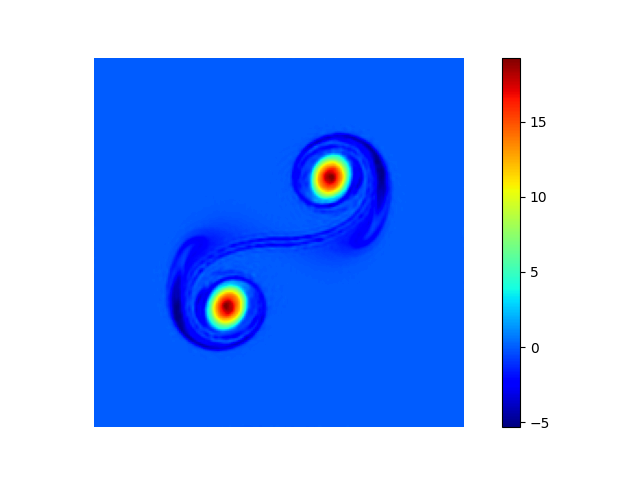}
  \end{minipage} \\[-6pt]
  \textbf{Ours \newline $5929\sim 6946$ particles \newline ($132.3$ KB)} & 
  \begin{minipage}{0.8\textwidth}\negvspace
    \formattedgraphics{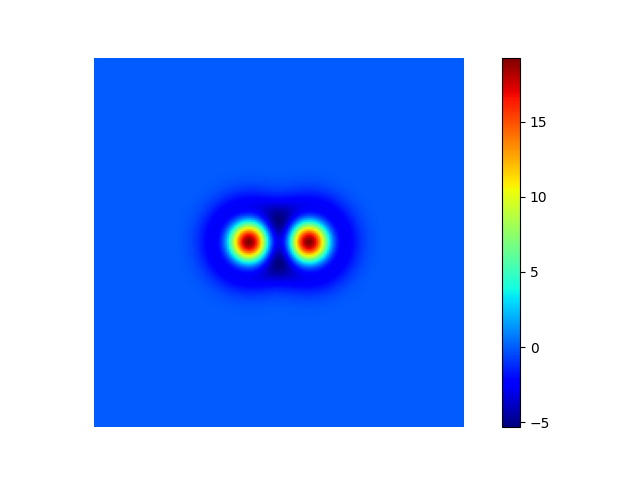}
    \negspace
    \formattedgraphics{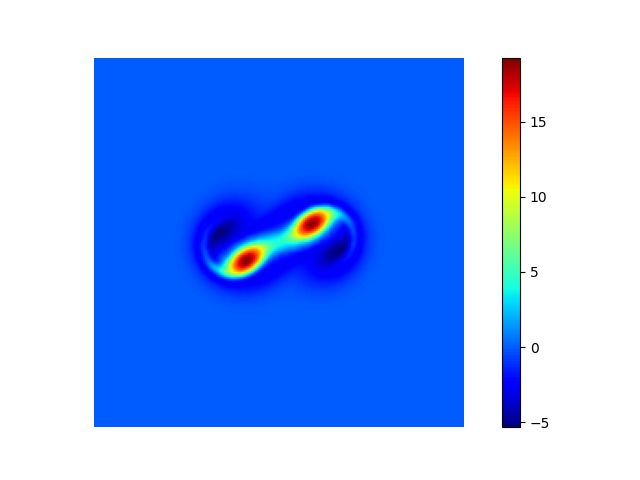}
    \negspace
    \formattedgraphics{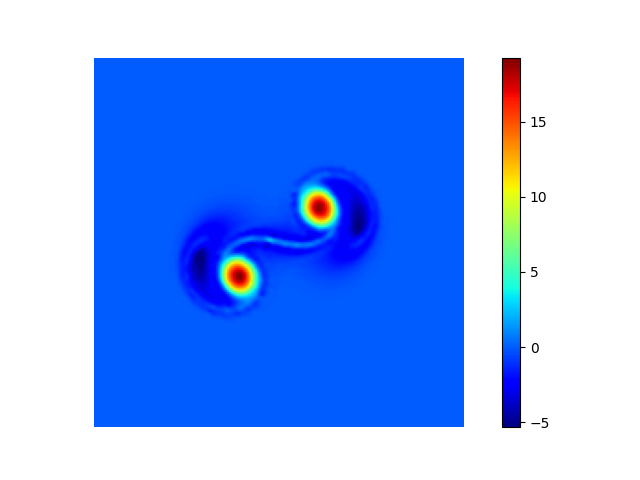}
    \negspace
    \formattedgraphics{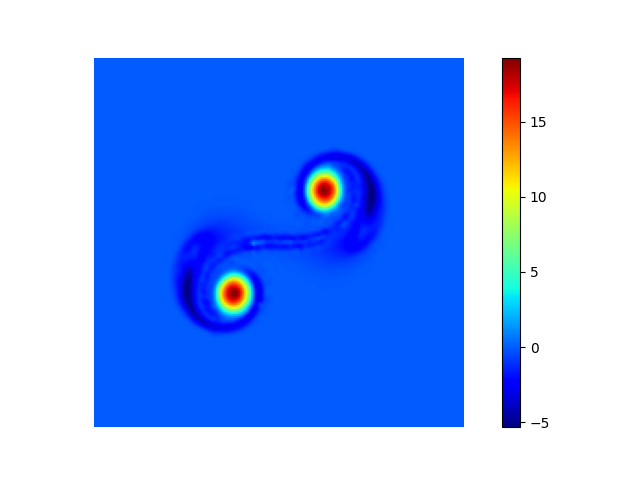}
    \negspace
    \formattedgraphicslst{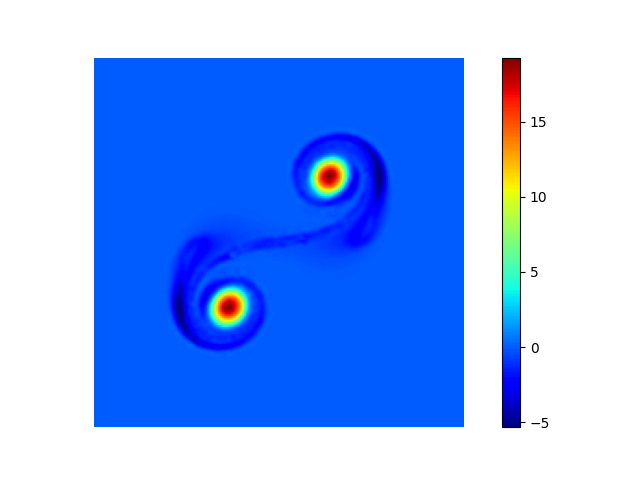}
  \end{minipage} \\[-6pt]
  \textbf{Ground truth: \newline vortex-in-cell \newline $512\times 512$ \newline ($2$ MB)} & 
  \begin{minipage}{0.8\textwidth}\negvspace
    \formattedgraphics{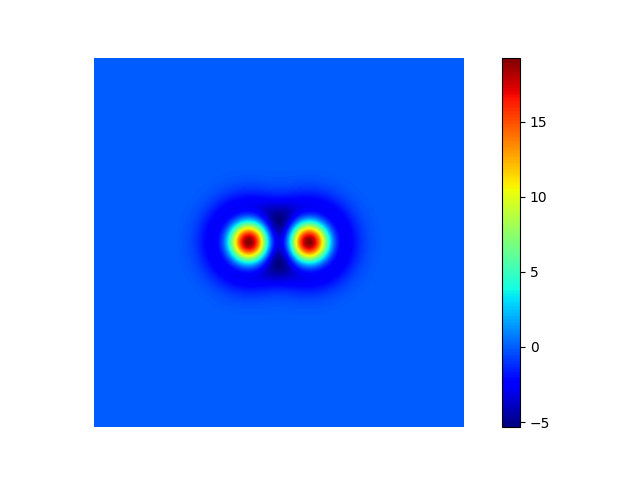}
    \negspace
    \formattedgraphics{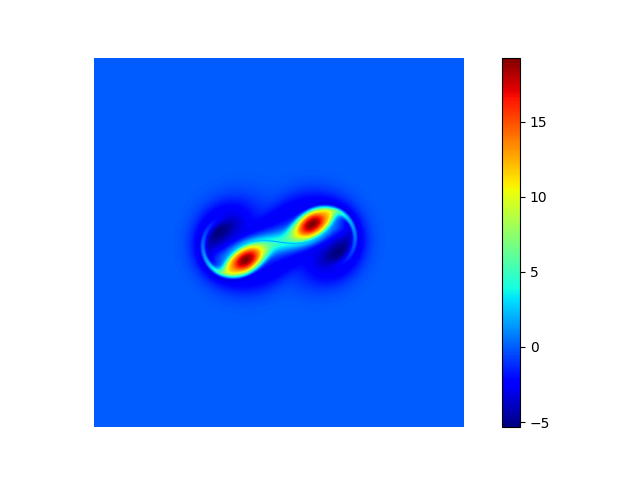}
    \negspace
    \formattedgraphics{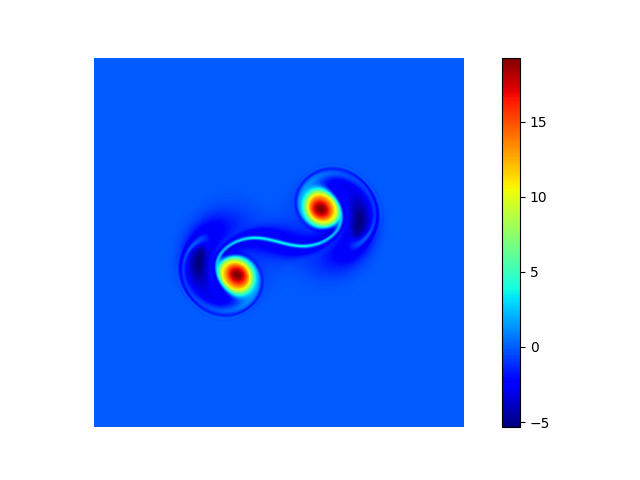}
    \negspace
    \formattedgraphics{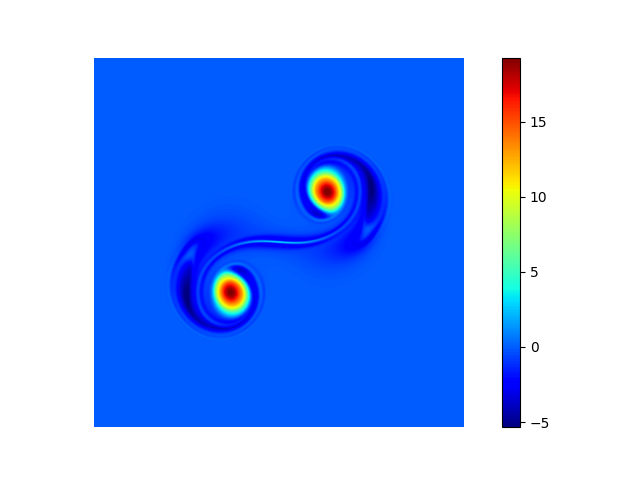}
    \negspace
    \formattedgraphicslst{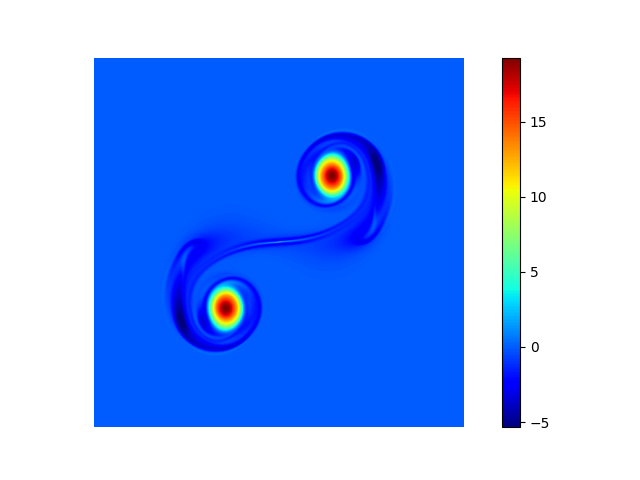}
  \end{minipage} \\[-6pt]
\end{tabular}
\caption{\footnotesize We compare our method with traditional Eulerian grid and different memory-efficient spatial representations INSR and GSR on the Taylor vortex example. Enclosed by parenthesis on the first column is the average file size of the according representation of the velocity field. Images show the vorticity field at frame 0, 100, 200, 300, and 399, respectively.}
\label{fig:taylor_vortex-full}
\end{figure*}

\paragraph{Efficacy of the DDFK}
We validate the efficacy of the DDFK by comparing our method with a traditional Eulerian solver, INSR and GSR on the Taylor vortex example, while use the simulation by vortex-in-cell method as the ground truth. Full results are shown in Figure~\ref{fig:taylor_vortex-full}. Our method manifests comparable ability on preserving vortices and details as the GSR, with significantly less noise on the background than INSR, indicating better stability of our method.
As illustrated in the figure, our method achieves vortex preservation and detail representation capabilities that are equally outstanding to those of GSR, with both methods accurately capturing the intrinsic structure and evolutionary dynamics of the Taylor vortex without significant degradation.
Despite sharing comparable detail preservation with INSR, our approach exhibits substantially less background noise while maintaining vortex magnitudes more precisely, which reflects superior stability of our DDFK solver.
Moreover, with far fewer degrees of freedom, our method captures the same flow details as the Eulerian solver while achieving better vortex preservation, which fully demonstrates its strong spatial adaptivity.

\paragraph{Validation of Zero-Divergence Preservation}
Two experiments are conducted to validate DDFK ability to preserve zero-divergence of the velocity fields throughout the simulation.
In the first example, a piston is moving downwards at a constant speed inside a closed vessel as shown in Figure~\ref{fig:teaser}, where the no-slip condition is applied on the piston and free-slip on  the vessel. The vessel is a square with its center at $(0,0)$ and a side length of $10$. The piston is a rectangle with a width of $8$ and a height of $0.5$, which is initially centered at $(0,2)$ and moves downward at a speed of $0.5$. Denote the segment of the line containing the upper edge of the piston that lies within the container as the set $\mathcal S_\mathrm{upper}$, the upper edge of the piston as $\mathcal E_\mathrm{upper}$, define the flowrate on curve $C$ as the integral of the $y$-component of the velocity field along $C$, denoted as $F_C$. If the fluid is incompressible,
\begin{equation}
    F_{\mathcal E_\mathrm{upper}}+F_{\mathcal S_\mathrm{upper}\backslash\mathcal E_\mathrm{upper}}=F_{\mathcal S_\mathrm{upper}}=0
\end{equation}
holds at any given time. Note that $\mathcal S_\mathrm{upper}\backslash\mathcal E_\mathrm{upper}$ is marked as the orange lines in Figure~\ref{fig:teaser}. Given the speed and width of the piston, we can calculate $F_{\mathcal E_\mathrm{upper}}=-0.5\times 8=-4$, hence the ground truth of $F_{\mathcal S_\mathrm{upper}\backslash\mathcal E_\mathrm{upper}}$ is $4$. We estimate $F_{\mathcal S_\mathrm{upper}\backslash\mathcal E_\mathrm{upper}}$ in the simulations using DFKs and GSR by calculating the average of the $y$-component of the velocity on uniformly sampled $400$ points on $\mathcal S_\mathrm{upper}\backslash\mathcal E_\mathrm{upper}$ times the length $\vert\mathcal S_\mathrm{upper}\backslash\mathcal E_\mathrm{upper}\vert$. As shown in the diagram of Figure~\ref{fig:teaser}, our solver DDFK maintains a flowrate near the ground truth throughout the entire simulation, while the flowrate of GSR is much lower. This also causes the vortices on both sides of the piston to surge higher with our method. In addition, as can be seen from the streamlines in Figure~\ref{fig:teaser} and the divergence field in Figure~, GSR failed to maintain a divergence-free field in the optimization based on the physics-informed divergence loss, whereas our method perfectly addressed this issue with the DFKs.

\begin{figure}[htbp]
\includegraphics[width=\linewidth]{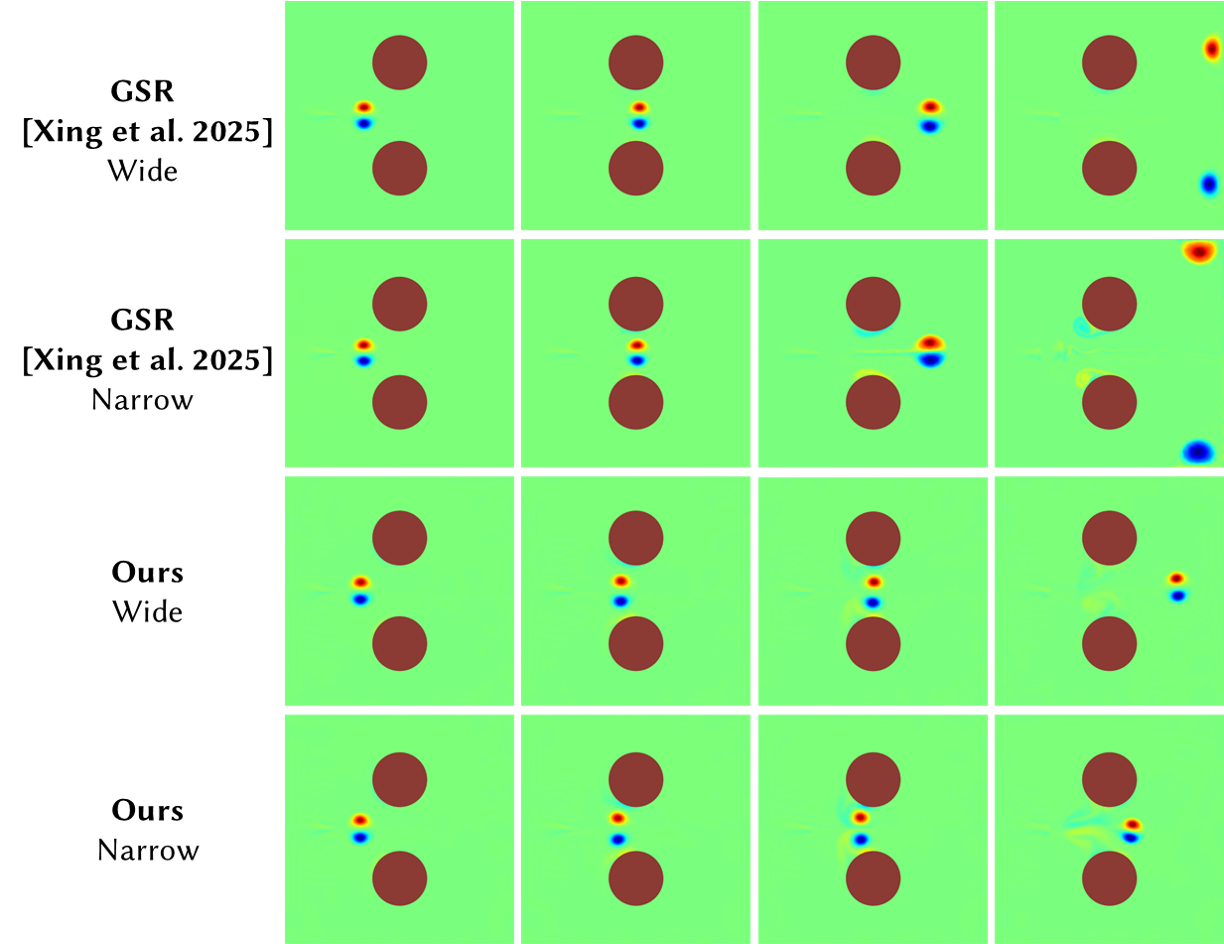}
\caption{We compare our method with Gaussian Fluids (GSR) \cite{xing2025gaussian-fluids} using vortices pass example of wide and narrow gaps. The wide setting
The positions of the two spherical obstacles in the wide and narrow
setting are $(0.5, 0.27) , (0.5, 0.73)$ and $(0.5, 0.285) , (0.5, 0.715)$. The radius of all the obstacles are $0.1174$.The figures from left to right are
showing frames 200, 336, 520 and 879, respectively.}
\label{fig:vortices_pass}
\end{figure}

In the second example, we compared two methods on the case where a pair of vortices pass through a gap between two spherical obstacles with a no-slip boundary as in Figure \ref{fig:vortices_pass}, where correctly modeling the harmonic component is key for successful traversal. We tested two scenarios with different spacing between the obstacles. Both methods can successfully allow the vortices to pass through the obstacles, but GSR's vortices pass through narrow obstacles faster than wide ones, and the vortex size also incorrectly increases. This occurs because the obstacles make it difficult for GSR's velocity field to maintain divergence-free conditions, while our method can naturally preserve divergence-free properties, thus correctly capturing the obstructive effect of the obstacles on vortex advancement - the vortices remain longer when passing through narrower gaps.

\begin{figure*}
    \centering
\includegraphics[width=.8\linewidth]{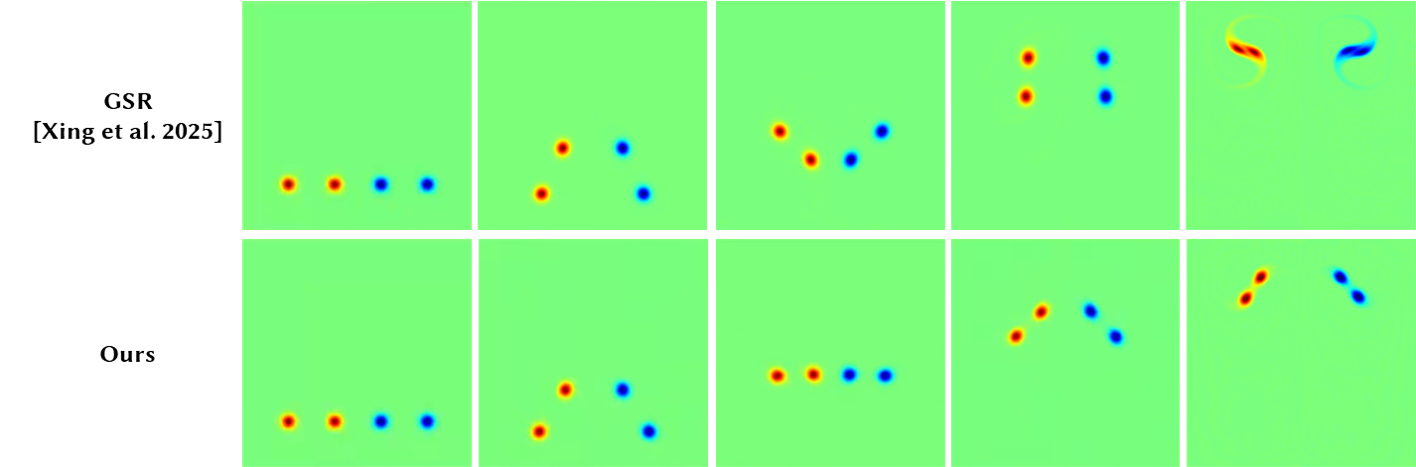} 
\vspace{-.5em}
\caption{Compare between our method and Gaussian Fluids (GSR) \cite{xing2025gaussian-fluids}. The images from left to right are
simulation results of Leapfrog 2D at frames 0, 165, 456, 1050 and 1500, respectively.}
\label{fig:leapfrog-2d}    
\end{figure*}

\paragraph{Low Numerical Dissipation}
We validate the numerical accuracy of our method with the Leapfrog experiment, where two vortex pairs released from the bottom engage in leapfrog-like rotation, moving upward. As shown in Figure~\ref{fig:leapfrog-2d}, comparing to Gaussian fluids in which vortex pairs merge before the fourth half-turn, ours can maintain the vortex separation and structure after the fifth half-turn. It shows the lower numerical dissipation of our method than Gaussian fluids. 

\begin{figure*}[htbp]
    \centering
    \newcommand{\formattedgraphics}[1]{\includegraphics[trim=100 50 230 50,clip,width=\columnwidth]{#1}}
    \begin{subfigure}{0.31\textwidth}
        \centering
        \formattedgraphics{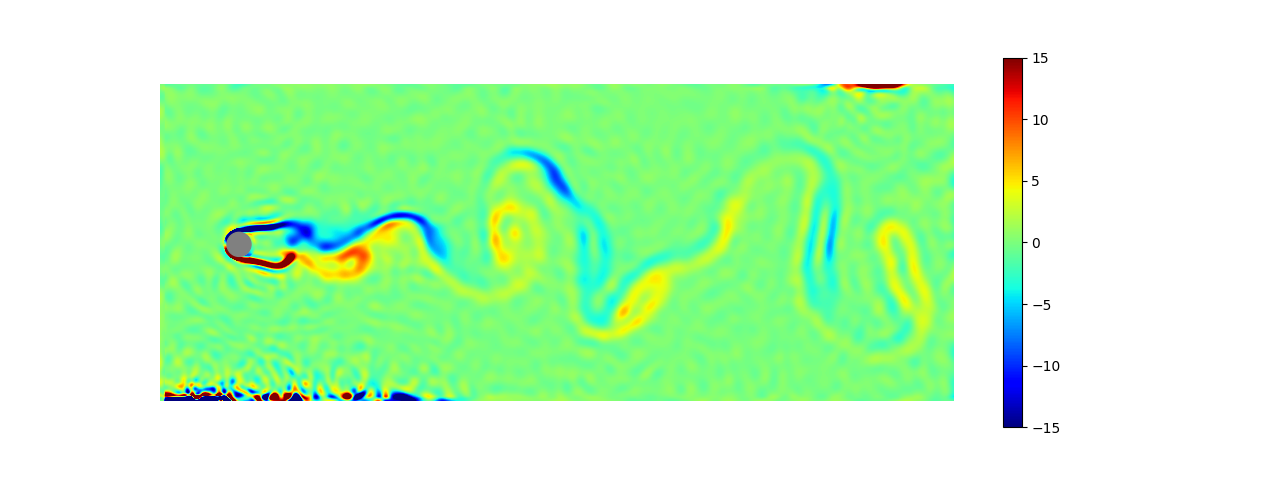}\vspace{-18pt}
        \caption{NMC}
        \label{fig:karman-NMC}
    \end{subfigure} ~
    \begin{subfigure}{0.31\textwidth}
        \centering
        \formattedgraphics{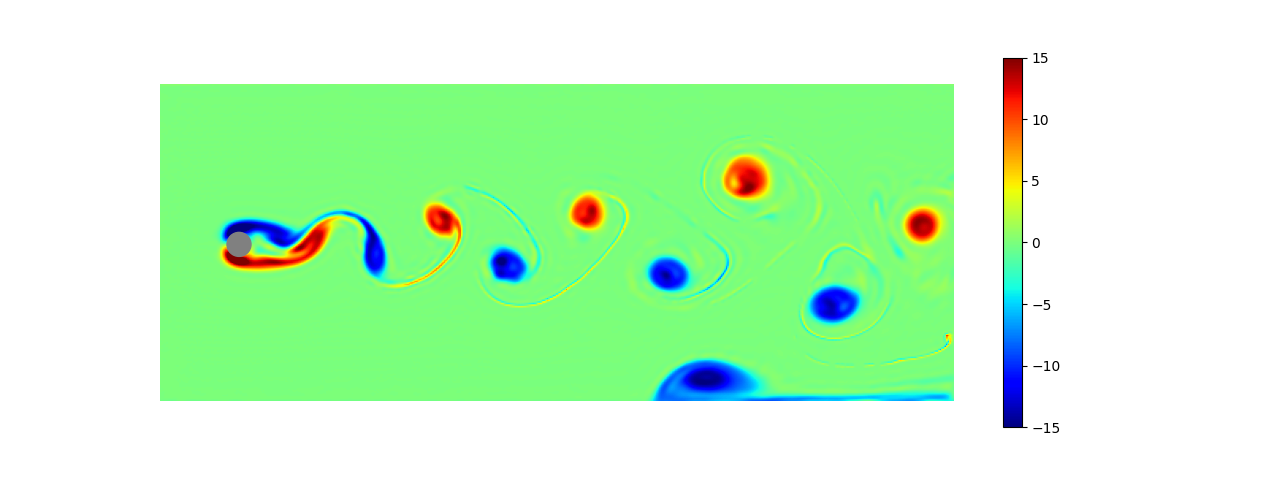}\vspace{-18pt}
        \caption{GSR}
        \label{fig:karman-ours}
    \end{subfigure} ~
    \begin{subfigure}{0.31\textwidth}
        \centering
        \formattedgraphics{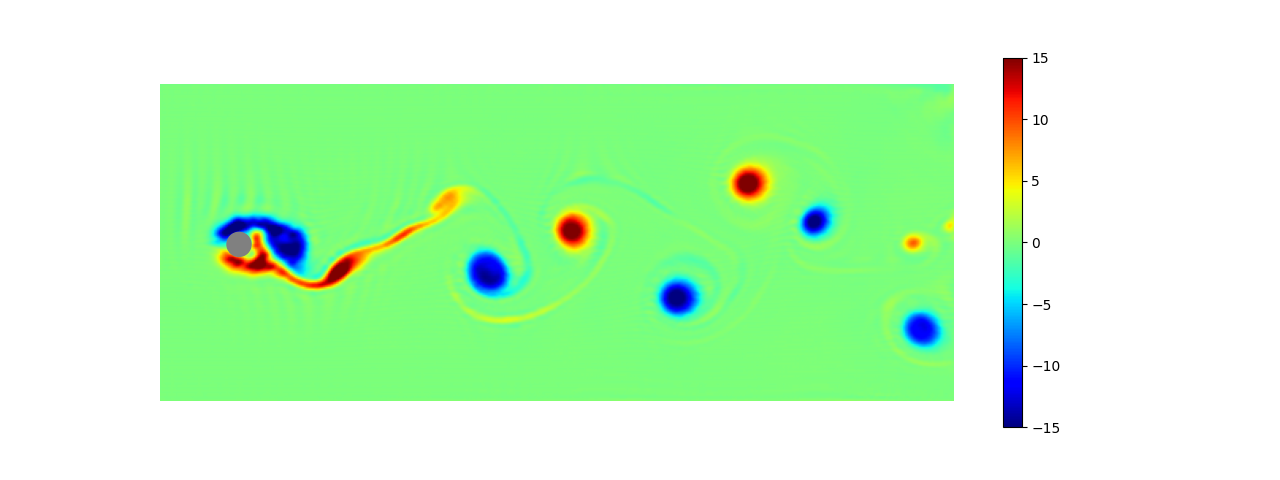}\vspace{-18pt}
        \caption{DDFK (Ours)}
        \label{fig:karman-SF+R}
    \end{subfigure}
    \vspace{-6pt}
    \caption{\footnotesize Karman vortex street example by Neural Monte Carlo (NMC), Gaussian fluids (GSR), and our method (DDFK). The sub-figures display the vorticity fields of frame 152, 199 and 199 of the simulation results, respectively.}\vspace{-8pt}
    \label{fig:karman}
\end{figure*}

\subsection{More Examples}

\paragraph{Karman Vortex Street}
In this example, fluid flows in from the left side of the domain, passes a cylindrical solid obstacle, forming alternating vortex street behind. As shown in Figure~\ref{fig:karman}, free-slip condition is applied on the upper and lower boundary of the domain, while no-slip condition is applied on the cylindrical obstacle. Meanwhile, free-slip condition with $f_\mathrm b$ set to the velocity of inflow is applied on the left and right boundary of the domain to keep the fluid flowing. We place extra particles outside of the left boundary of the domain in initialization and reinitialization steps to prevent particle deficiency due to advection. Results in Figure~\ref{fig:karman} indicates both our method and Gaussian fluids have better stability than neural Monte Carlo method based on INSR, while our simulation results manifests higher Reynolds number with irregular vortex street behind the obstacle.

\subsection{Ablation Tests}
\label{sec:ablation}

\begin{figure*}
    \centering
    \includegraphics[width=.7\linewidth]{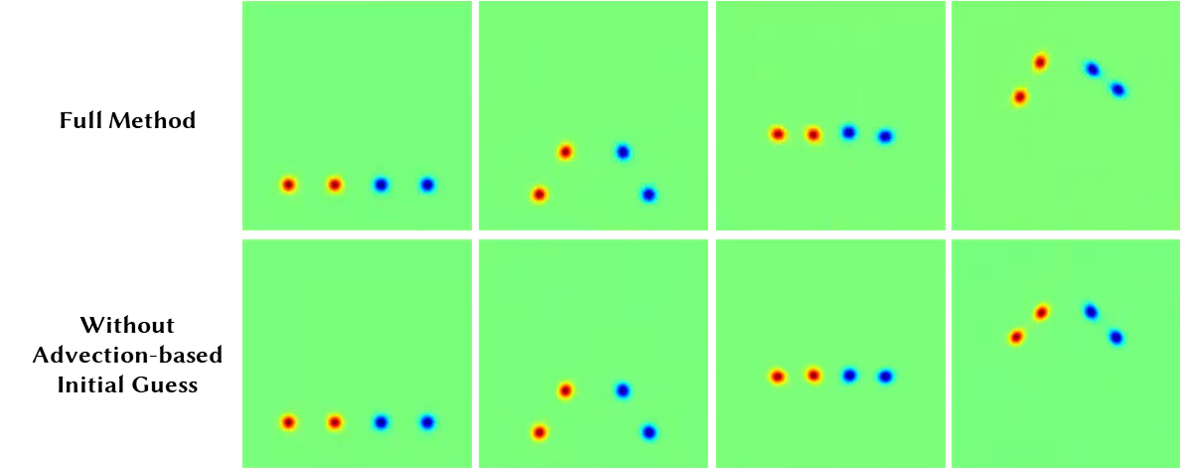} 
\vspace{-.5em}
\caption{Ablation test on the advection-based initial guess. The images from left to right are
simulation results of Leapfrog 2D at frames 0, 150, 533 and 1000, respectively.}
\label{fig:leapfrog-2d-ablation}    
\end{figure*}

\paragraph{Advection-Based Initial Guess} 
The Initial Guess improve the accuracy of our method. As show in Figure~\ref{fig:leapfrog-2d-ablation}, we compares our method with and without the initial guess under the leapfrog 2D example. the result produced by our method without initial guess becomes asymmetric after long-term advection, while the full method's result remains symmetric and maintains the rotation of the vortex pair for a longer duration. Numerically, we can also observe that although the vorticity field without advection-based initial guess has a lower loss before optimization, the vorticity field with initial guess achieves a lower loss after optimization. This demonstrates that initial guess enables our method to perform more accurate advection.

\paragraph{Reinitialization}

\begin{figure}[htbp]
    \centering
    \begin{subfigure}{0.45\columnwidth}
        \centering
        \includegraphics[width=\textwidth]{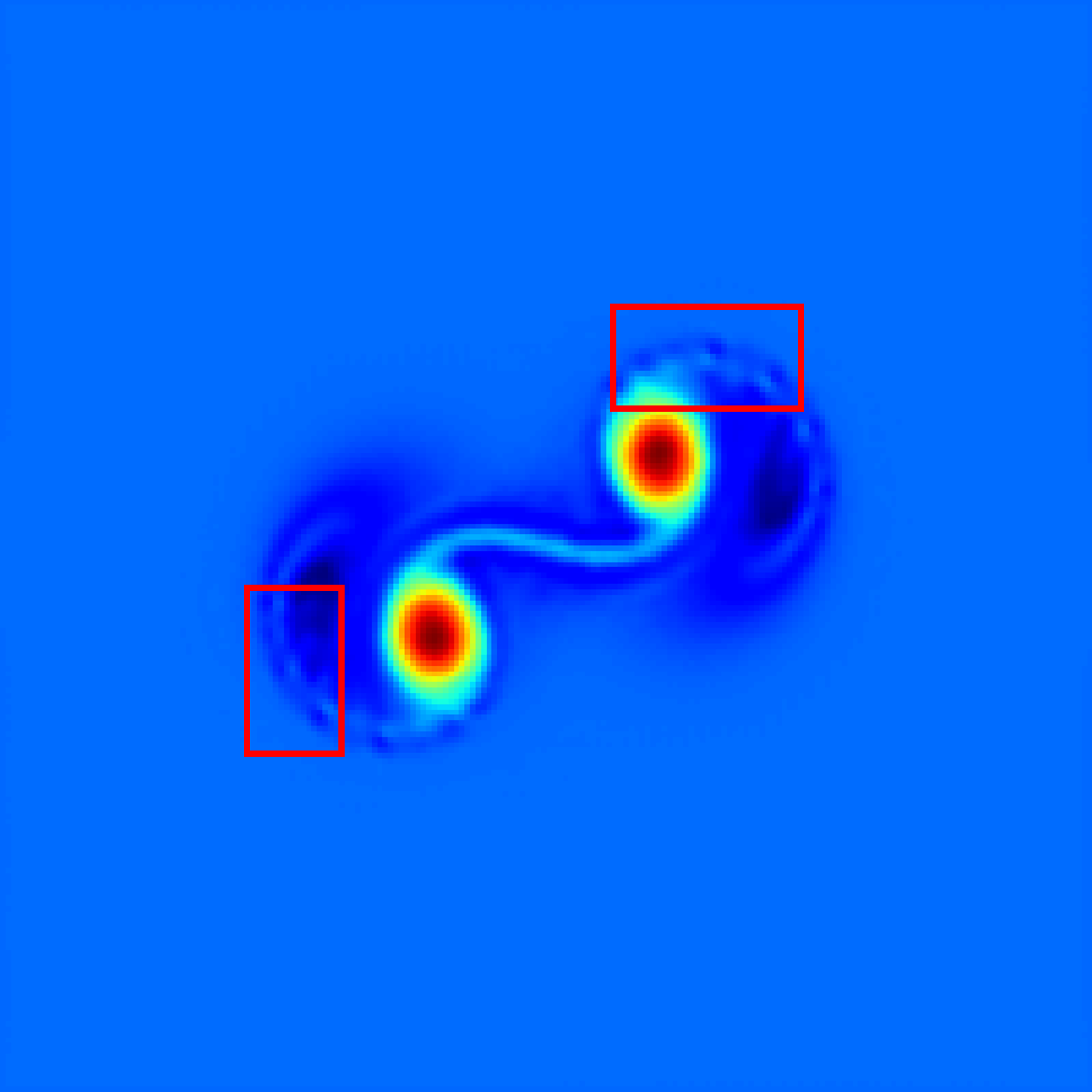}
        \caption{Without reinitialization}
        \label{fig:taylor_vortex-ablation-noreinit}
    \end{subfigure} ~
    \begin{subfigure}{0.45\columnwidth}
        \centering
        \includegraphics[width=\textwidth]{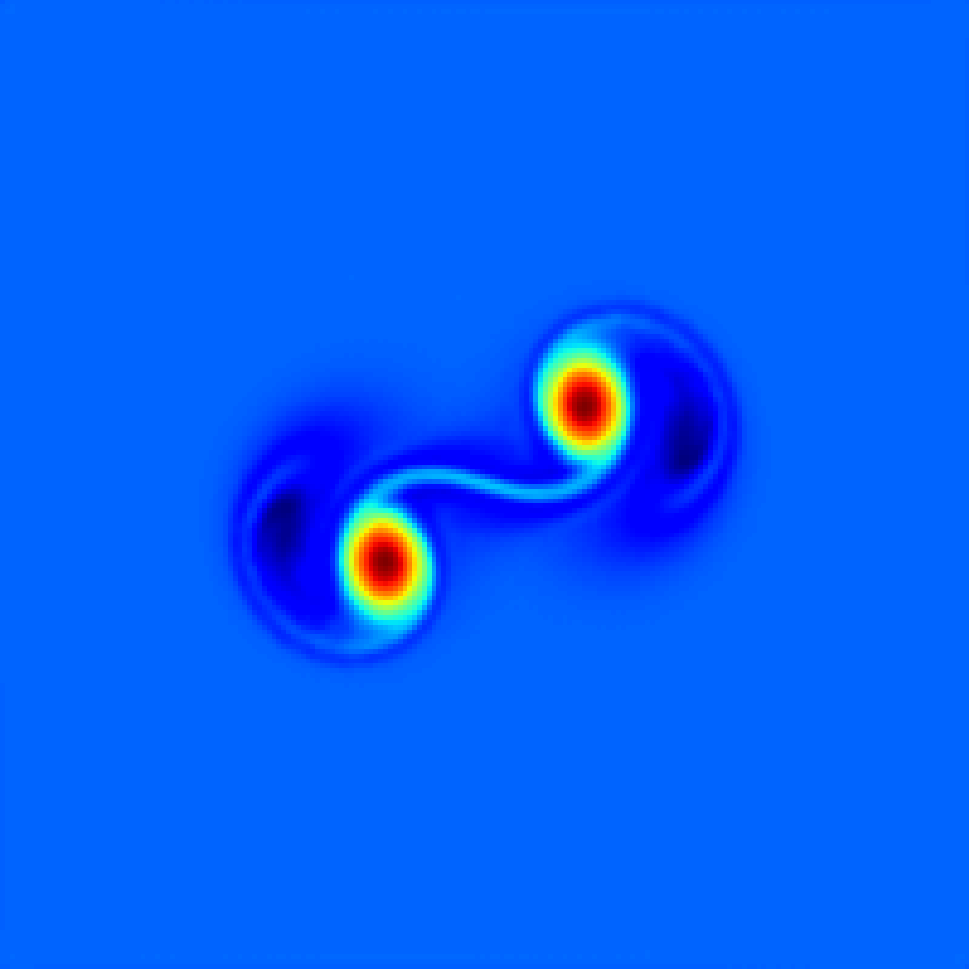}
        \caption{Full method}
        \label{fig:taylor_vortex-ablation-full}
    \end{subfigure} \\
    \begin{subfigure}{0.45\columnwidth}
        \centering
        \includegraphics[width=\textwidth]{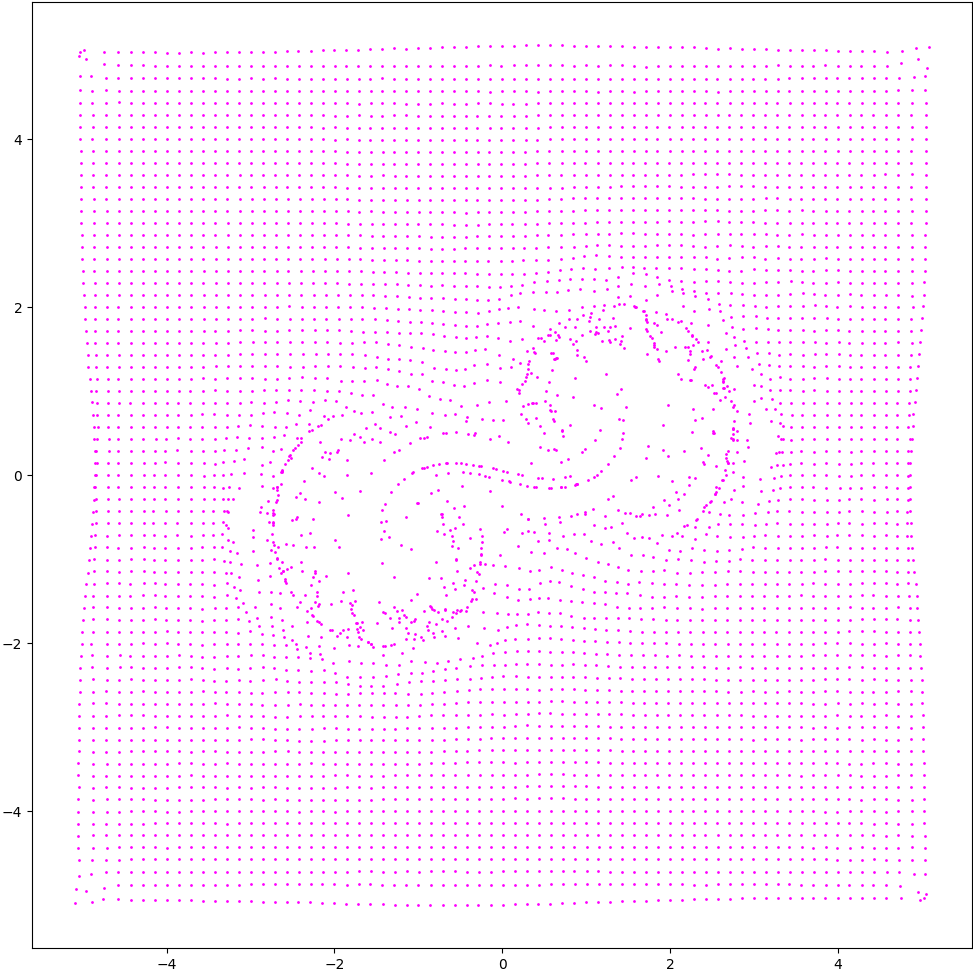}
        \caption{$\{\boldsymbol p_i\}$ without reinitialization}
        \label{fig:taylor_vortex-ablation-noreinit-positions}
    \end{subfigure} ~
    \begin{subfigure}{0.45\columnwidth}
        \centering
        \includegraphics[width=\textwidth]{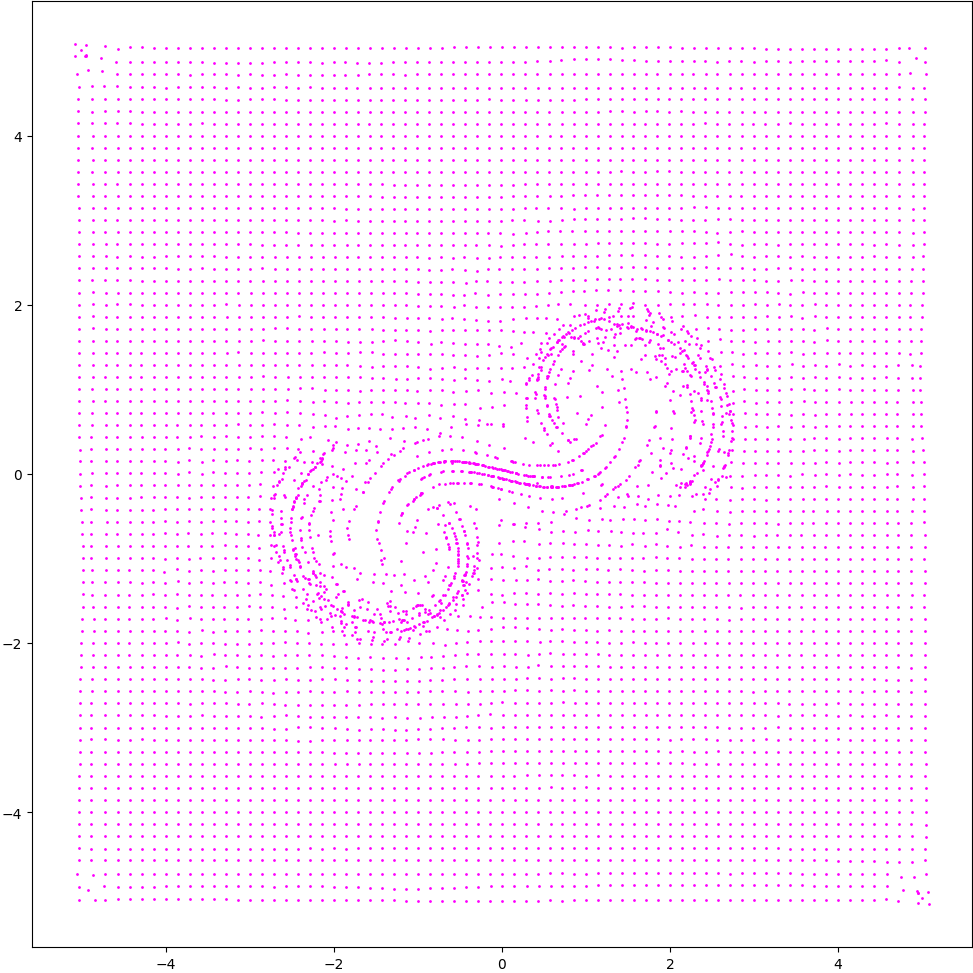}
        \caption{$\{\boldsymbol p_i\}$ in full method}
        \label{fig:taylor_vortex-ablation-full-positions}
    \end{subfigure}
    \caption{Ablation study on reinitialization. We show the simulation results and particle positions at frame 180.}
    \label{fig:taylor_vortex-ablation}
\end{figure}

Figure~\ref{fig:taylor_vortex-ablation} compares the method without reinitialization with the full method. The former manifests noisy artifacts near the tails of the vortices, as marked in the red boxes in Figure~\ref{fig:taylor_vortex-ablation-noreinit}, due to particle deficiency as shown in Figure~\ref{fig:taylor_vortex-ablation-noreinit-positions}. The reinitialization process effectively fill particles into these under-fitting areas, resulting in a cleaner simulation.

\section{Conclusion}

In this paper, we propose a grid-free solver based on dynamic divergence-free kernels for incompressible flow. Our method perfectly preserves zero-divergence, resolving the main issue on optimizing the physics-informed divergence loss with other memory-efficient spatial representations like INSRs and GSR. Meanwhile, the DDFK also have the advantages of INSRs and GSR over traditional discretizations: spatial adaptivity, continuous differentiability and vorticity fidelity. Moreover, our method can stably simulate fluid phenomena in a long time horizon, with generalization across scenarios with intricate vortices and diverse boundaries.

\clearpage

\clearpage
\bibliographystyle{ACM-Reference-Format}
\bibliography{refs}

\end{document}